\documentclass[aps,prb,floatfix,twocolumn,superscriptaddress,showpacs,amsmath,amssymb]{revtex4}
\usepackage{graphicx}
\usepackage{epsfig} 
\usepackage{color}

\allowdisplaybreaks

\newcommand{\be}{\begin{equation}}
\newcommand{\ee}{\end{equation}}
\newcommand{\bea}{\begin{eqnarray}}
\newcommand{\eea}{\end{eqnarray}}
\newcommand{\ba}{\begin{eqnarray*}}
\newcommand{\ea}{\end{eqnarray*}}
\newcommand{\dagga}{{\phantom{\dagger}}}
\newcommand{\bR}{\mathbf{R}}
\newcommand{\bQ}{\mathbf{Q}}

\newcommand{\bk}{\mathbf{k}}

\newcommand{\up}{\uparrow}
\newcommand{\down}{\downarrow}

\newcommand{\eqn}[1]{(\ref{#1})}
\newcommand{\m}[1]{\mathcal{#1}}

\newcommand{\ket}[1]{\mid #1 \rangle}
\newcommand{\bra}[1]{\langle #1 \mid}

\begin{document}
\title{Non-equilibrium gap-collapse near a first-order Mott transition } 

\author{Matteo Sandri}
\affiliation{
International School for Advanced Studies (SISSA), and CRS Democritos, CNR-INFM, - Via Bonomea 265, I-34136 Trieste, Italy} 
\author{Michele Fabrizio}
\affiliation{
International School for Advanced Studies (SISSA), and CRS Democritos, CNR-INFM, - Via Bonomea 265, I-34136 Trieste, Italy} 

\date{\today} 

\begin{abstract}
We study the non-equilibrium dynamics of a simple model for V$_2$O$_3$ that consists of a quarter-filled Hubbard model for two 
orbitals that are split by a weak crystal field. Peculiarities of this model are: (1) a Mott insulator whose gap corresponds to transferring an electron from the occupied lower orbital to the empty upper one, rather than from the lower to the upper Hubbard sub-bands; (2) a Mott transition generically of first order even at zero temperature. We simulate by means of time-dependent Gutzwiller approximation the evolution within the insulating phase of an initial state endowed by a non-equilibrium population of electrons in the upper orbital and holes in the lower one. We find that the excess population may lead, above a threshold, to a gap-collapse and drive the insulator into the metastable metallic phase within the coexistence region around the Mott transition. This result foresees a non-thermal pathway to revert a Mott insulator into 
a metal. Even though this physical scenario is uncovered in a very specific toy-model, we argue it might 
apply to other Mott insulating materials that share similar features. 

\end{abstract}

\maketitle

\section{Introduction}
Mott insulators are potentially promising candidates that might enable scalability below the size 
of conventional semiconductor solid state devices.\cite{Newns1997} In fact, Mott insulators can typically 
revert to metals, e.g. under pressure, suddenly releasing the large amount of conduction electrons that were 
earlier Mott localized. Therefore one may envisage that an external stimulus, like a voltage bias 
or an intense optical pulse, could eventually drive a Mott insulator into a metal with a very large carrier concentration. 

Experimental attempts performed so far are indeed encouraging, see e.g. Refs. \onlinecite{Iwasa-2012} and \onlinecite{Cario-2013}. On the contrary, theoretical calculations in the simplest model for a Mott insulator, namely the half-filled single-band Hubbard model, are not equally promising. For instance, the simulated time evolution of a photo-excited Mott insulator, with holes in the lower Hubbard band and electrons in the upper one, shows that the injected energy effectively heats the system, which relaxes to a thermal 
steady-state the slower the stronger the interaction.~\cite{Werner-PRB-2011,Werner-PRL-2013,Prelovsek-PRL-2013}  In other words, the Mott-Hubbard sub-bands persist and simply spectral weight is transferred from the lower to the upper, just as if temperature rises, though 
small deviations from the expected thermal behavior are observed.~\cite{Werner-PRB-2011,Freericks-PRL-2013} Moreover, theoretical simulations of the dielectric breakdown of a single-band Mott insulator with gap $E_\text{Gap}$ 
in the presence of a static electric field $\text{E}$ point towards a conventional Landau-Zener mechanism, 
i.e. tunneling between 
lower and upped Hubbard bands over a distance $\sim E_\text{Gap}/\text{E}$, not dissimilar  
to conventional band insulators.~\cite{Oka-PRL-2003,Okamoto-PRB-2007,
Okamoto-PRL-2008,Oka&Aoki-PRB-2010,Martin-PRL-2010,Martin-PRB-2013} 

These results are evidently a bit disappointing, all the more so since they are not even in full accordance with experiments.~\cite{Cario-2013,Nagaosa-PRB-2008} A simple escape route, which we shall follow here,  is to abandon the half-filled single-band Hubbard model as the prototypical model to describe dielectric breakdown in real Mott insulators. Indeed, we note that the Mott insulating materials where the dielectric breakdown has been experimentally observed so far, at least to our knowledge, all have
a charge gap that is either of charge-transfer origin, like NiO,\cite{NiO-APL-2008}
Cu$_2$O,\cite{Cu2O-APL-2010} or cuprates,~\cite{Newns1998,Goldman-PRB}, or it is  
an inter-band gap between Mott-localized occupied orbitals and unoccupied ones, all sharing the same 
atomic $d$-character,  
like VO$_2$,\cite{VO2-JAP-2009} or V$_2$O$_3$.~\cite{Cario-2013} In none of these cases the charge gap is therefore the genuine Mott-Hubbard gap that refers to the same element and same orbital. The natural question is therefore if and how this feature affects the off-equilibrium response to external perturbations that could drive those materials metallic. 

We shall try here to elucidate this question in a very simple model that we originally introduced to 
reproduce qualitatively the physics of V$_2$O$_3$,~\cite{Sandri_finT}
and which does describe a Mott insulator with a charge-gap between occupied and unoccupied orbitals. Specifically, we shall study by means of the time-dependent Gutzwiller approximation the temporal evolution 
of a non-equilibrium initial state characterized by an excess population of particle-hole excitations, where 
holes lie in the lower Hubbard band of the occupied orbital while particles sit in the unoccupied 
conduction orbital. Our aim is to ascertain whether the non-equilibrium initial condition only causes heating, 
hence leaves well defined and separated conduction and valence bands, although the former slightly occupied 
 and the latter slightly emptied, or rather the system evolves into a non-thermal and possibly metallic phase.

The paper is organized as follows. In Sec.~\ref{Sec2} we introduce the model and discuss its equilibrium phase diagram. In Sec.~\ref{The model out-of-equilibrium} the peculiar equilibrium properties are invoked  
to envisage a non-equilibrium pathway able to drive the Mott insulator into a metastable metal. 
We briefly sketch the time-dependent Gutzwiller approximation in Sec. \ref{sec4} and study the outcome of the aforementioned non-equilibrium protocol in Sec. \ref{sec5}, both in the paramagnetic and antiferromagnetic sector. Finally, Sec. \ref{Conclusions} is devoted to conclusions.

\section{The model at equilibrium}
\label{Sec2}
The model we shall consider is a two-band Hubbard model that we originally designed to capture the main physics of vanadium sesquioxide V$_2$O$_3$.~\cite{Sandri_finT} We believe, however, that this model is simple enough to provide information of more general validity, even beyond the physics pertaining to 
V$_2$O$_3$. 

Specifically, we assume on each site of a two-dimensional square lattice two orbitals that are split by a crystal field. Each site is occupied on average by a single electron. In addition, we include a Hubbard repulsion that penalizes  configurations where the on-site occupation is different from one, and intra- and inter-orbital hopping elements between nearest neighbor sites. We do not include any Coulomb exchange splitting, which would implement Hund's rules, because it is ineffective for configurations with a single electrons thus we expect it would only add unnecessary complications. The Hamiltonian is 
\bea
&&\mathcal{H} = \sum_{a=1}^2\sum_{\bk\sigma}\, 
\epsilon_{\bk}\, c^\dagger_{a\bk\sigma}c^\dagga_{a\bk\sigma} 
+ \sum_{\bk\sigma}\,\gamma_\bk\, \big(c^\dagger_{1\bk\sigma}c^\dagga_{2\bk\sigma}+H.c.\big)
\nonumber\\
&& + \sum_i\bigg[
-\Delta\,\big(n_{1i}-n_{2i}\big) + \frac{U}{2}\,\big(n_{1i}+n_{2i}-1\big)^2\bigg],\label{Ham}
\eea
where $a=1,2$ labels the two orbitals, $\epsilon_\bk = -2t\big(\cos k_x + \cos k_y\big)$ is 
the standard nearest neighbor tight-binding energy, $U$ parametrizes the on-site repulsion and 
$\Delta>0$ the crystal field splitting. We include an inter-orbital hopping 
$\gamma_\bk = -4t'\,\sin k_x\sin k_y$, where we hereafter set $t'=0.3\,t$, with a symmetry such that the 
local single-particle density matrix remains diagonal in the orbital indices 1 and 2. Such choice is made 
because we want to require that the occupation of each orbital is not a conserved quantity and yet that 
both orbitals are irreducible representations of the crystal field symmetry. 

\begin{figure}[hbt]
\includegraphics[width=7.5cm]{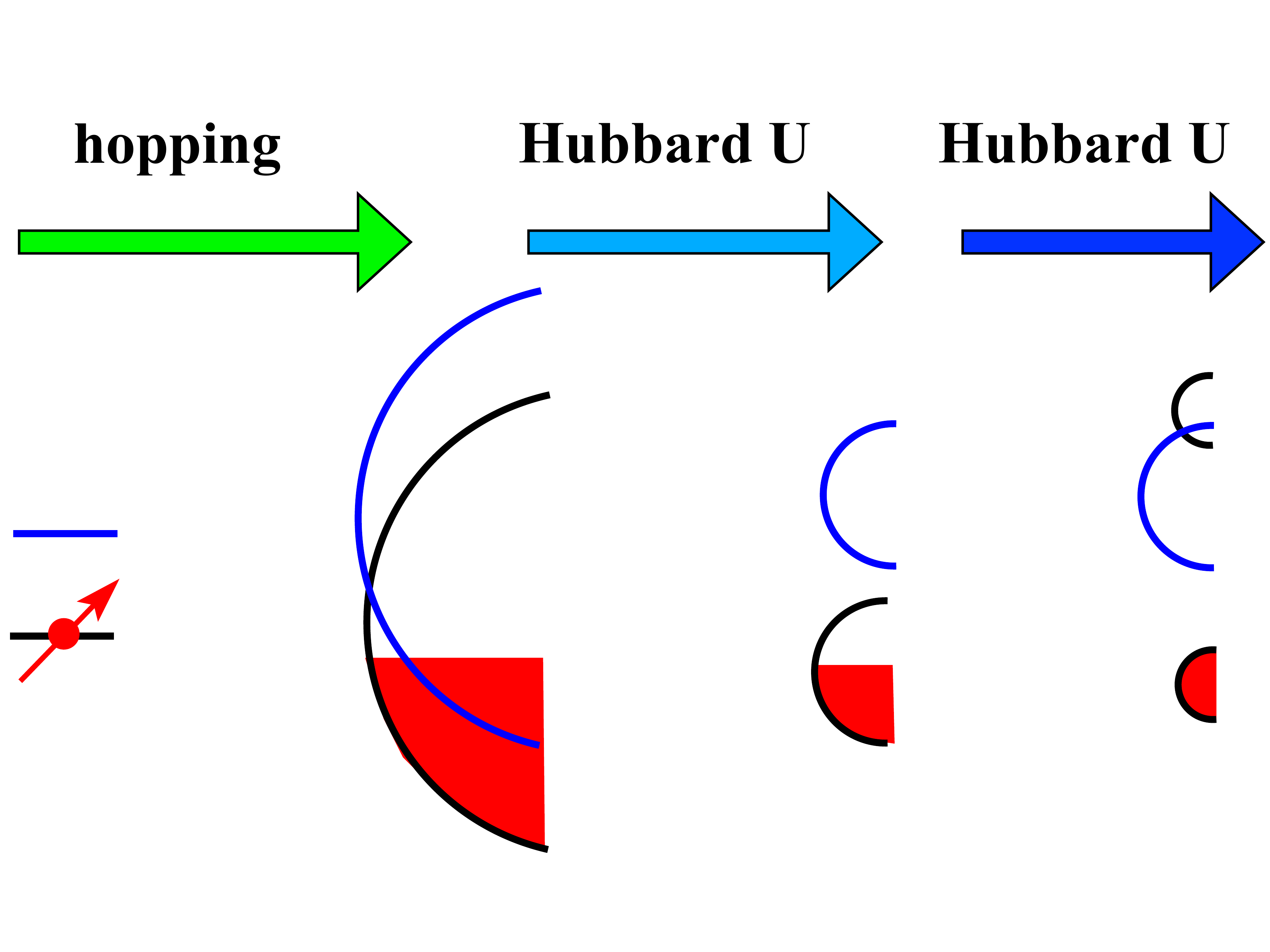}
\caption{(Color online) Phases displayed by the model Eq. \eqn{Ham}. From the left: in the isolated site the electron occupies the lowest orbital. When sizable hopping is switched on, the model describes a quarter-filled two-band metal. If we now turn on interaction, the coherent bandwidth shrinks and meanwhile the crystal field splitting increases, so that the higher band gradually empties, until only the lowest one remains populated, actually half-filled. A further increase of $U$ will then drive this band across a Mott metal-to-insulator transition, with lower and upper Hubbard bands.}
\label{picture}
\end{figure}
\begin{figure}[!h]
\includegraphics[scale=0.32]{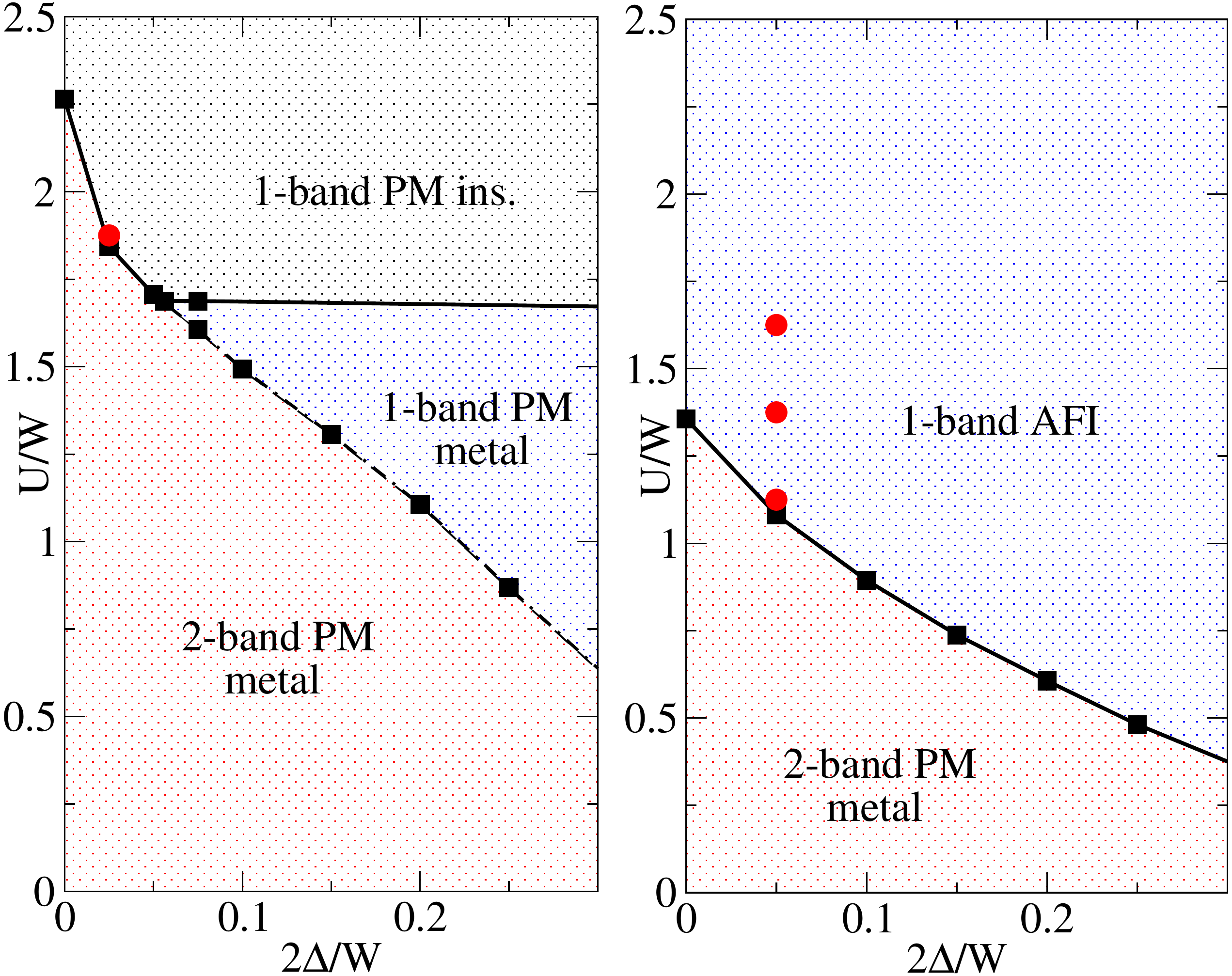}
\caption{(Color online) $T=0$ phase diagram for the model Hamiltonian (\ref{Ham}) obtained by the Gutzwiller approximation in the paramagnetic  sector (left panel) and antiferromagnetic one (right panel). PM stands for paramagnetic while AFI for antiferromagnetic insulator. The red dots represent the values of $\Delta$ and $U$ that we shall consider in the dynamics. The energy unit is $W=8t$, where $t$ is the orbital-diagonal hopping. (From Ref. \onlinecite{Sandri_finT}). }
\label{fig:phaseT0}
\end{figure}
\begin{figure}[!h]
\includegraphics[scale=0.32]{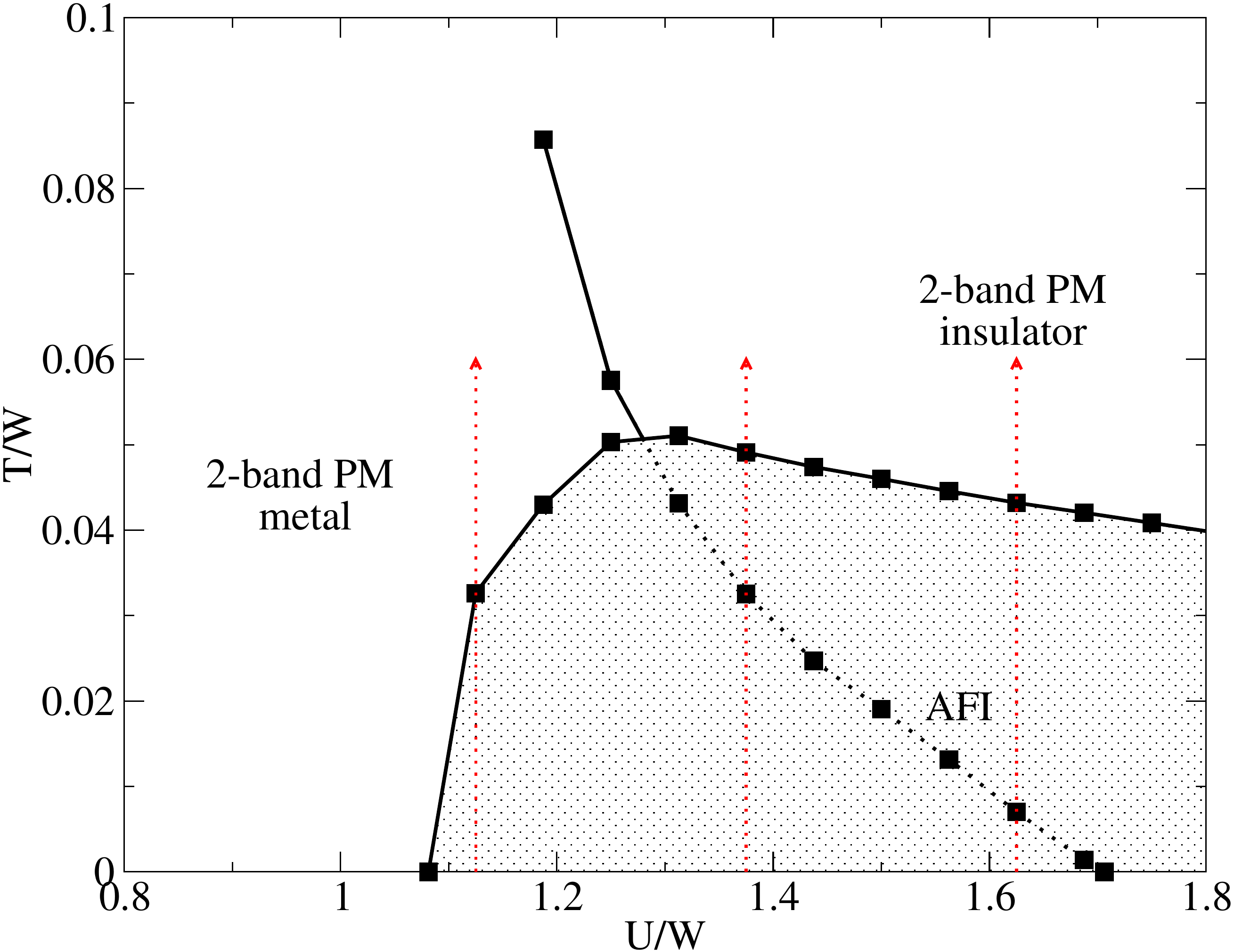}
\caption{(Color online) Finite temperature phase diagram of the model Eq.~(\ref{Ham}) obtained by the Gutzwiller approximation at $\Delta =0.025W$.  Within the stability dome of the antiferromagnetic insulator (AFI), we indicate the first-order line that would separate the metal from the Mott insulator should magnetism be not allowed. The red arrows show the values of $U$ considered in the dynamics.
(From Ref. \onlinecite{Sandri_finT}.)}
\label{fig:phaseFT}
\end{figure}

Let us first briefly discuss the possible phases displayed by the model Eq.~\eqn{Ham} in connection with the relative strengths of the Hamiltonian parameters, which we sketch in Fig. \ref{picture} and 
locate in the zero-temperature phase-diagrams shown in Fig.~\ref{fig:phaseT0}, where magnetism is not allowed, and in Fig.~\ref{fig:phaseFT} in the more realistic magnetic case.
In the atomic limit, 
leftmost side in Fig.~\ref{picture}, the single electron occupies the lowest orbital. When the hopping is turned on, the orbitals broaden into two bands that we shall assume hereafter overlap so much that both are occupied; the uncorrelated model thus describes a quarter-filled two-band metal. Switching on the interaction $U$ brings two distinct effects. On one side the repulsion between occupied and unoccupied states effectively increases the crystal field splitting; a phenomenon that can be well described also within any independent particle scheme, as e.g. mean-field theory. Such approximate schemes are however unable to capture another important interaction effect that is the bandwidth shrinking, which reduces the band overlap hence enhances further the strength of the crystal field. 
The crystal field splitting growth and the bandwidth shrinking gradually empty the higher band and eventually leave only the lowest band populated, actually half-filled, second to last drawing 
in Fig.~\ref{picture}. A further increase of $U$ can then drive the lowest half-filled band towards a Mott insulating phases, likely accompanied by the emergence of magnetism.

The actual phase diagram at $T=0$ that we previously obtained by the Gutzwiller approximation, see 
Ref.~\onlinecite{Sandri_finT}, is shown in Fig.~\ref{fig:phaseT0} as function of the crystal field $\Delta$ and Hubbard $U$, using $W=8t$ as energy unit . The left panel is the case in which we artificially force the solution to stay paramagnetic (PM). It agrees qualitatively and to some extent also quantitatively well with more reliable dynamical mean-field theory (DMFT) calculations by Poteryaev, Ferrero, Georges and Parcollet.~\cite{Ferrero-2band}   

On the contrary, there are not DMFT results to compare with in the physical case in which we allow the solution to spontaneously order magnetically, in our case antiferromagnetism is actually favored. This corresponds to the right panel in 
Fig. \ref{fig:phaseT0}. Here only a first order transition from a two-band paramagnetic metal to a 
one-band antiferromagnetic Mott insulator is found by the Gutzwiller approximation. Although both phase diagrams in Fig.~\ref{fig:phaseT0} span a wide range of crystal field values, we shall here concentrate only in the small-$\Delta$ region, see red dots in the figure, which we believe is more representative of V$_2$O$_3$. 

Two aspects that characterize the model Eq.~\eqn{Ham} at $T=0$ 
and small $\Delta$ are worth to be highlighted 
in comparison with the single-band Hubbard model. In the Mott insulator, while the lower band is split as usual into a lower and upper Hubbard sub-bands, the unoccupied band actually undresses from correlations, so that the lowest charge excitation corresponds to transferring an electron from the lower Hubbard band to the lesser correlated valence band.~\cite{Ferrero-2band,Sandri_finT} Moreover, when magnetism is allowed, the one-band metal phase is predicted to disappear at zero temperature leaving a direct {\sl first-order} phase transition between a paramagnetic two-band metal and an antiferromagnetic Mott insulator.~\cite{Sandri_finT} In other words, on the insulating side close to the Mott transition a metastable paramagnetic metal phase is expected to exist even at zero temperature, an interesting feature that foreshadows the possibility to stabilize such a phase under non-equilibrium conditions, for instance by an external bias as in the phenomenological model  proposed in Ref.~\onlinecite{Cario-2013} to explain experimental data.

To complete our discussion on the equilibrium properties of the model Eq.~\eqn{Ham}, we show in 
Fig.~\ref{fig:phaseFT} the $T\not=0$ phase diagram at $\Delta=0.025\,W$.~\cite{Sandri_finT}
It comprises three different phases: a two-band paramagnetic metal at small $U$, a two-band paramagnetic Mott insulator 
at large $U$ and $T$, and finally an antiferromagnetic Mott insulator at large $U$ and low $T$. In this case of finite temperature, even the higher band gets thermally populated, 
hence we generically denote the phases as two-bands. Without magnetism, the paramagnetic metal is 
separated from the paramagnetic Mott insulator by a first order line, shown in the figure.  In our specific two-band model at quarter filling, this first order line is not entirely covered by magnetism unlike in the single-band case. Indeed, there is still a segment that emerges from the magnetic dome and ends up into a second order critical point, closely resembling the phase diagram of V$_2$O$_3$ that was actually the target material 
this model was designed for. It is important for the following analysis to highlight that also the N\'eel transition is here first order. Therefore, when the coexistence region close to the N\'eel  transition overlaps with the coexistence region between paramagnetic metal and paramagnetic insulator, 
all three distinct phases exist, although only one is thermodynamically stable while the other two are metastable.  
We finally mention that the qualitative features 
of the finite-temperature phase diagram in Fig.~\ref{fig:phaseFT}, specifically the order of the transitions and 
the existence of a first-order line above the magnetic region, agree with more reliable DMFT calculations performed at the same value of crystal field $\Delta=0.025\,W$.~\cite{Sandri_finT}

\section{The model out-of-equilibrium}
\label{The model out-of-equilibrium}

We shall now study the model Eq.~\eqn{Ham} in out-of-equilibrium conditions. The guiding idea 
is very simple. We mentioned already that the Hubbard $U$ introduces a repulsion between occupied and 
unoccupied states that effectively enhances the crystal field. For instance,  the center of gravity of the  higher band ``2" with respect to the lower band ``1" increases within mean-field from 
$2\Delta$ to 
\be
2\Delta_\text{eff} = 2\Delta + \frac{U}{2}\,\big(n_1-n_2\big)\equiv 
2\Delta + \frac{U}{2}\,m
,\label{MF-Delta}
\ee
where $n_1$ and $n_2$ are the average occupations of each orbital, and 
$m=n_1-n_2$ is hereafter defined as the {\sl orbital polarization}. This effect in turns anticipates the 
Mott transition, which thus occurs at lower $U$ the higher the population imbalance $m$. 
Vice versa, we can also imagine that a sudden reduction of $m$ from the $m\simeq 1$ value in the Mott insulating phase, induced for instance by an intense light pulse, may launch an avalanche process -- the reduced $m$ makes
$\Delta_\text{eff}$ smaller, which in turn decreases $m$ further and so on -- thus pushing temporarily the system in the stability region of the two-band metal.  

\begin{figure}[!h]
\includegraphics[scale=0.32]{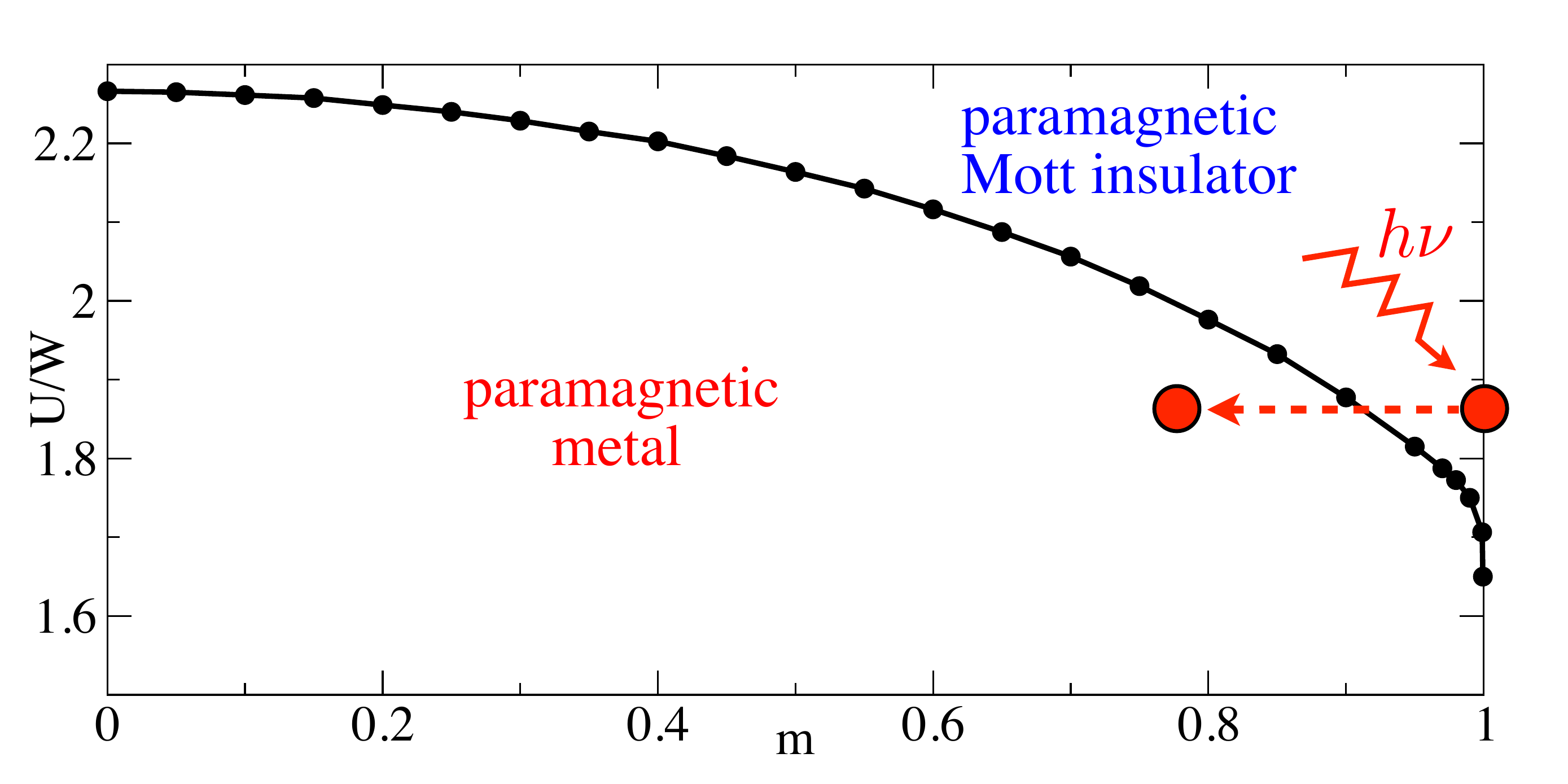}
\caption{(Color online) Critical $U$ for the Mott transition at fixed value of the orbital polarization 
 $m=n_1-n_2$. We also show that hypothetical path of the system under the action of a light pulse of energy $h\nu$.}
\label{MIT.vs.m}
\end{figure}

\begin{figure}[!h]
\includegraphics[scale=0.32]{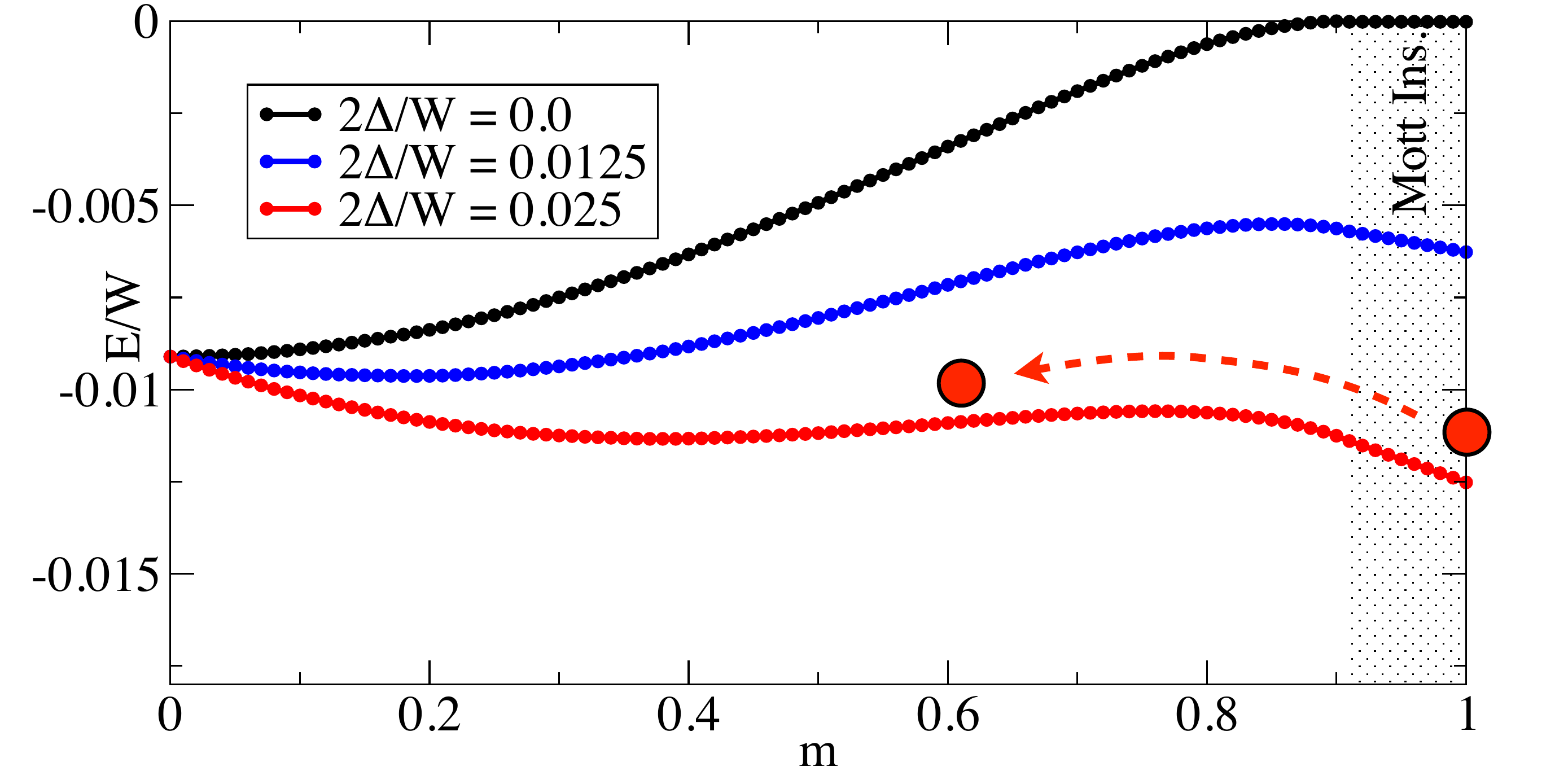}
\caption{(Color online) Total energy density at $U=1.875 W$ as function of the 
orbital polarization $m$ and for different crystal field splittings $2\Delta$ in the paramagnetic sector. The gray area corresponds to the insulating phase. We observe the appearance of two minima at small $\Delta\not = 0$. We also show the hypothetical out-of-equilibrium process that drives a system initially in the stable Mott insulating phase towards the metastable metal one.}
\label{EnevsM_PM}
\end{figure}

\begin{figure}[!h]
\includegraphics[scale=0.32]{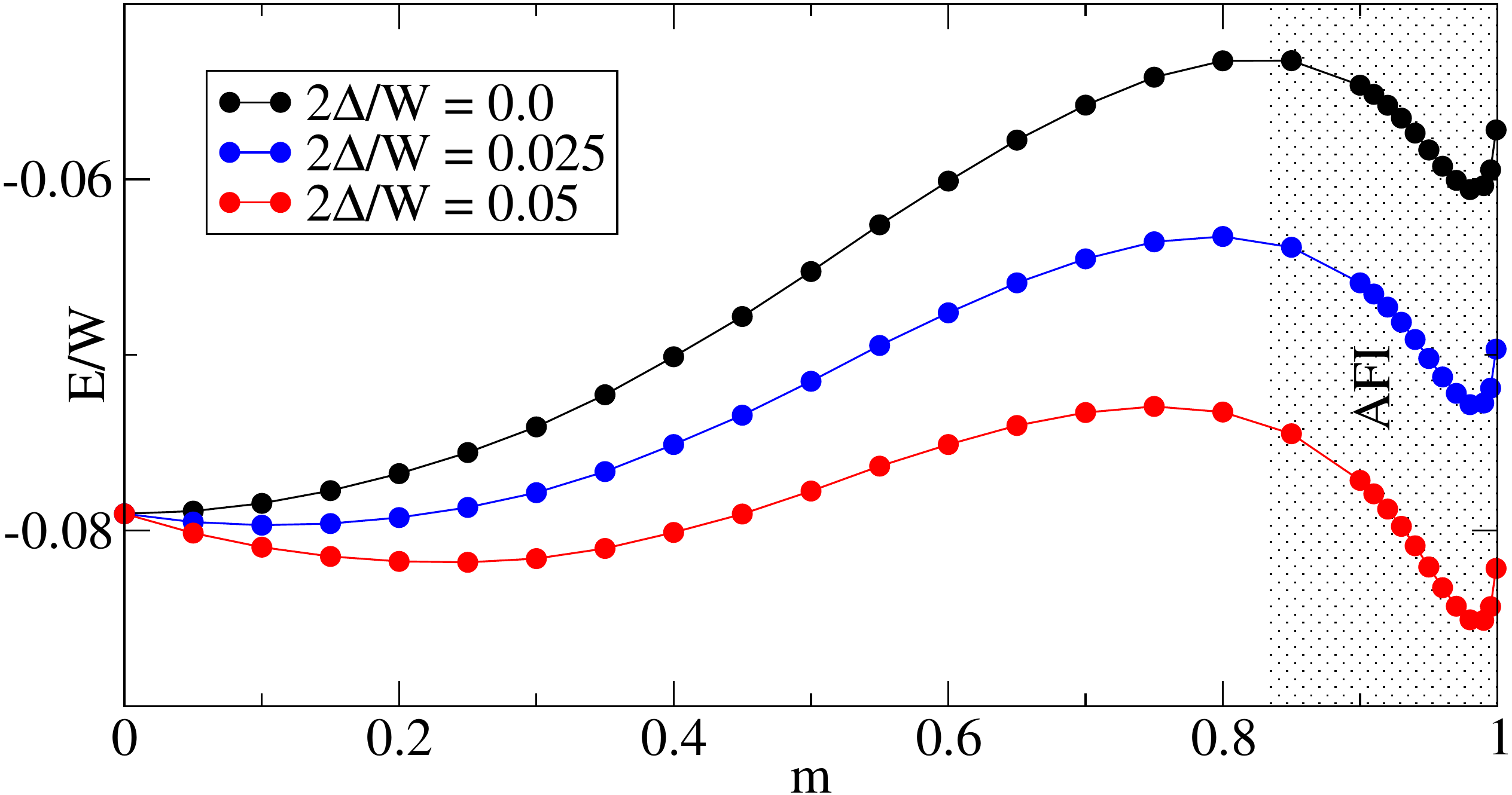}
\caption{(Color online) Same as Fig. \ref{EnevsM_PM} but allowing for magnetism. In this case the gray area corresponds to an antiferromagnetic Mott insulator (AFI).}
\label{EnevsM_AFM}
\end{figure}

This is illustrated schematically in Fig. \ref{MIT.vs.m}, where we show 
the $T=0$ phase diagram as function of $U$ and of the orbital polarization $m$ instead of its conjugate variable $\Delta$, as obtained by 
 the Gutzwiller approximation~\cite{Sandri_finT, Hvar} in the paramagnetic sector. 
In the same figure we also sketch the following hypothetical experiment. The system is initially in the $T=0$ $m\simeq1$ Mott insulating phase, with the lower orbital occupied and Mott localized whereas the upper one empty, see Fig.~\ref{picture}. We then imagine that a light pulse suddenly transfers a certain amount of electrons from the lower to the upper orbital, thus reducing the orbital polarization $m$, as shown in Fig.~\ref{MIT.vs.m}. The system is thus temporarily pushed in stability region of the two-band metal phase. In the subsequent evolution, the system could either equilibrate back to the insulator with a thermally reduced orbital polarization, or it could remain trapped into a metastable metallic phase with overlapping bands.  
The latter event is indeed not unlikely, as we may evince by inspection of the energy as function 
of $m$ at fixed $U$ and for various $\Delta$'s. In Figs.~\ref{EnevsM_PM} and \ref{EnevsM_AFM} we plot for instance such energy at $U=1.875 W$ in the paramagnetic sector and $U=1.125W$ in the magnetic one, respectively. We observe that there is a whole range of crystal field values where two minima coexist, one at $m\simeq 1$ corresponding to a single band Mott insulator, and another at smaller $m$ that describes a two-band metal. 
The first order Mott transition that we mentioned earlier just corresponds to the energy crossing of these two minima. It follows that, on the Mott insulating side of the coexistence region,  one cannot exclude  the possibility that, by suddenly reducing the orbital polarization, the system could be indeed trapped in the metastable metallic solution, as sketched in both Figs.~\ref{EnevsM_PM} and \ref{EnevsM_AFM}. This is just the scenario we shall try to uncover. 

Before presenting our simulation of such a hypothetical out-of-equilibrium metallization, we 
caution about an important aspect that we will not be able to capture correctly. Since we are considering 
a system artificially driven into a metastable phase within a coexistence region 
around a first order phase transition, nucleation effects are expected to play a role. However, since we shall  
always consider a homogeneous state, we will be not able to describe nucleation  
both of the metastable metal phase and of the stable Mott insulating one. Therefore we do not expect to describe well long-time relaxation. This is made even worse by the fact that the approximate method we use to simulate the non-equilibrium dynamics does not account for all physical dissipative processes.

\section{The method: time-dependent Gutzwiller approximation}
\label{sec4}

In order to simulate the non-equilibrium dynamics we shall use the 
time-dependent Gutzwiller approximation~\cite{Schiro&Fabrizio_prl} (t-GA) that has been extensively discussed elsewhere. Here we briefly sketch the method in the formulation of Ref.~\onlinecite{Hvar}, which the reader is referred to for more details.

The main idea of the time dependent Gutzwiller technique is to approximate the evolving wavefunction $\ket{\Psi(t)}$ in terms of a variational state whose dynamics is set by requiring the stationarity of the real time action 
\be \label{LAG}
\m{S}(t) = \int_0^t d\tau \; \bra{\Psi(\tau)}  i \partial_\tau - \m{H}(\tau) \ket{\Psi(\tau)}.
\ee
One introduces the following ansatz for the wavefunction~\cite{Schiro&Fabrizio_prl}
\be \label{GWF}
\ket{\Psi(t)} = \prod_{\bR} \m{P}_{\bR}(t) \;\ket {\Psi_0(t)},
\ee 
where $\ket {\Psi_0(t)} $ is a generic time-dependent variational Slater determinant, and $\m{P}_{\bR}(t)$ a local time-dependent variational operator.

In our model, although the inter-orbital hybridization $t'$ is finite, hence the two orbitals can mutually exchange electrons, still the local density matrix is diagonal by symmetry. It is thence convenient to use as local basis the Fock space
\be \label{BASE}
\ket{ \bR, \{n\} } = \prod_{a=1,2} \prod_{\sigma=\up,\down}(c^\dagger_{\bR a\sigma})^{n_{a\sigma}}\mid \!0\rangle, 
\ee
where $\{n\}=(n_1,n_2)$, and we shall denote the single-orbital states as $n_a=0$, if the orbital $a=1,2$ is empty, $n_a=\sigma$, with $\sigma=\up$ or $\down$, 
if it is occupied by a single spin-$\sigma$ electron, and finally 
$n_a=2$ if it is occupied by two electrons. 

The linear operator $\m{P}_{\bR}(t)$ can be parametrized in terms of a set of time dependent variational parameters $\Phi_{\bR \;\{n\}}(t)$ as 
\be
\m{P}_{\bR}(t) = \sum_{\{ n\}} \frac{\Phi_{\bR \; \{n\}}(t)}{\sqrt{P^{(0)}_{\bR \; \{n\} }(t) }}
\;  \ket {\bR, \{n\}}\langle {\bR, \{n\}} | ,
\ee
where
\be
P^{(0)}_{\bR \; \{n\} }(t) = \langle \Psi_0(t) \ket {\bR, \{n\}} \langle {\bR, \{n\}} \ket{\Psi_0(t)},
\ee
is the time-dependent occupation probability of the local state 
$\ket{n}$ in the uncorrelated Slater determinant. Within the Gutzwiller approximation, the occupation probability $P_{\bR \; \{n\} }(t)$ in the correlated wavefunction $\ket{\Psi(t)}$ 
turns out to coincide with the variational parameter 
$\left|\Phi_{\bR\;\{n\}}(t)\right|^2$. 

In Ref.~\onlinecite{Hvar} it was shown that the stationarity of (\ref {LAG}) amounts to solve a set of coupled differential equations that determine the evolution of the uncorrelated wavefunction $\ket{\Psi_0(t)}$ and of the variational parameters $\Phi_{\bR \;\{n\}}(t)$:
\bea 
& &i\, \partial_t \,\ket{\Psi_0(t)} = \m{H}_*\Big[\hat{\Phi}(t)\Big]  \ket{\Psi_0(t)}, \label{TE1}\\
& &i\, \partial_t \, \hat{\Phi}_{\bR}(t)  = \Big( \hat{U}_\bR + \hat{\Delta}_\bR \Big)\, \hat{\Phi}_{\bR}(t) 
\nonumber \\
&& \qquad \qquad \qquad + \, \langle \Psi_0(t) \mid\frac{\partial \m{H}_* [\hat{\Phi}(t)]}{\partial \hat{\Phi}_{\bR}(t)^\dagger } \mid \Psi_0(t)\rangle. \label{TE2}
\eea
With the notation $\hat{O}_\bR$ we indicate the matrix representation of the local operator $\mathcal{O}_\bR$ in the Fock basis (\ref{BASE}). $\m{H}_*\big[\hat{\Phi}(t)\big]$ is a non-interacting time-dependent Hamiltonian that depends parametrically on the variational parameters $\Phi_{\bR \;\{n\}}(t)$. Physically, Eq.~\eqn{TE1} describes the dynamics of the coherent quasiparticles whereas Eq.~\eqn{TE2} that of the incoherent atomic-like excitations, including the Hubbard bands. These two types of excitations are only coupled in a mean-field like fashion within the time-dependent Gutzwiller approximation, a weak point that makes the method unable to describe all dissipative processes of the real dynamical evolution.   

In the general case of a N\`eel order on a bipartite lattice,~\cite{Sandri_AFM} 
the effective quasiparticle Hamiltonian $\m{H}_*$ reads
\bea \label{RenHam}
\m{H}_* &=& \sum_{\bk,\sigma} \bigg\{ \sum_{a=1,2}\, \epsilon_\bk \Big[ \Re\left(Z_{a}(t)\right) \, c^\dagger_{\bk a\sigma}c^\dagga_{\bk a\sigma}\nonumber\\
&& \qquad \qquad \qquad - i\Im\left(Z_{a}(t)\right)\, c^\dagger_{\bk a\sigma}c^\dagga_{\bk+\bQ a\sigma} \Big]  \\
	&+& \gamma_\bk \Big[ Z^S_{\sigma}(t) \, 
	c^\dagger_{\bk 1\sigma}c^\dagga_{\bk 2\sigma} + 
	Z^A_{\sigma}(t)\,  c^\dagger_{\bk 1\sigma}c^\dagga_{\bk+\bQ 2\sigma}    + H.c. \Big ] \bigg\},
	\nonumber
\eea
with $\bQ = (\pi, \pi)$ the magnetic wave vector. We have defined the following quantities:
\bea \label{Zetas}
Z_{a}(t) &=& R^*_{a \sigma}(t) R^\dagga_{a -\sigma}(t), \\
2\,Z^S_{\sigma}(t) &=& R^*_{1 \sigma}(t) R^\dagga_{2 \sigma}(t) + R^*_{1 -\sigma}(t) R^\dagga_{2 -\sigma}(t), \nonumber \\
2\, Z^A_{\sigma}(t) &=& R^*_{1 \sigma}(t) R^\dagga_{2 \sigma}(t) - R^*_{1 -\sigma}(t) R^\dagga_{2 -\sigma}(t), \nonumber
\eea
where $R_{a\sigma}$ are the renormalization parameters that occur in the Gutzwiller variational approach, 
whose meaning is that the Fermi operator  $c^\dagga_{\bR a\sigma}$ 
at site $\bR$ belonging to the sublattice $A$ has a coherent quasiparticle content that, after projection, reads  
\be
\mathcal{P}_\bR(t)^\dagger \,c^\dagga_{\bR a\sigma}\,\mathcal{P}_\bR(t)^\dagga 
\rightarrow R^\dagga_{A,a\sigma}(t)\,c^\dagga_{\bR a\sigma} \equiv R^\dagga_{a\sigma}(t)\,c^\dagga_{\bR a\sigma},\label{def:R}
\ee
while the quasiparticle content of the operators on sublattice $B$ is obtained simply by recalling that 
\be
R_{B,a\sigma}(t) = R_{A,a -\sigma}(t) \equiv R_{a-\sigma}(t).
\ee
A similar relation holds for the variational parameters 
$\Phi_{\bR\;\{n\}}(t)$, too, so that we shall hereafter drop the label $\bR$ and refer always to a generic site in sublattice $A$. 

The parameters $R_{a\sigma}(t)$ thus play an important role as they determine whether or not coherent quasiparticles exist. At equilibrium $R_{a\sigma}(t)=R_{a\sigma}$ can always be chosen real and its vanishing signals the 
onset of the paramagnetic Mott insulator. Out-of-equilibrium but in the half-filled single-band model, $R_{a\sigma}(t)$ can still be chosen real and oscillates in time around a well defined mean value,~\cite{Schiro&Fabrizio_prl} whose vanishing identifies a dynamical Mott transition. In the present two-band model, $R_{a\sigma}(t)$ is unavoidably complex because its phase has to generate the effective one-body potential for the quasiparticles that induces the crystal field splitting as well as the staggered Zeeman splitting. Indeed, if we write
\be
R_{a\sigma}(t) = \text{e}^{i\phi_{a\sigma}(t)}\;\rho_{a\sigma}(t),
\label{R-complex}
\ee
with $\rho_{a\sigma}(t)$ real, we can absorb the phase $\phi_{a\sigma}(t)$ into a unitary transformation, see Eq. \eqn{def:R},
\be
c^\dagga_{\bR a \sigma} \to \text{e}^{-i\phi_{a\sigma}(t)}\;
c^\dagga_{\bR a \sigma},
\ee
for $\bR$ in the $A$-sublattice, and instead 
\be
c^\dagga_{\bR a \sigma} \to \text{e}^{-i\phi_{a-\sigma}(t)}\;
c^\dagga_{\bR a \sigma},
\ee 
for $\bR$ in the $B$-sublattice. After such unitary transformation the quasiparticle dynamics is controlled by a new Hamiltonian 
\bea
\widetilde{\m{H}}_*(t) &=& V_*(t) + \sum_{\bk,\sigma} \Bigg\{ \sum_{a=1,2}\, \epsilon_\bk \, \widetilde{Z}_{a}(t) \, c^\dagger_{\bk a\sigma}c^\dagga_{\bk a\sigma} \nonumber\\
&& \qquad + \gamma_\bk \bigg[ \widetilde{Z}^S_{\sigma}(t) \, 
	c^\dagger_{\bk 1\sigma}c^\dagga_{\bk 2\sigma} \label{H_*-vera} \\
	&& + 
	\widetilde{Z}^A_{\sigma}(t)\,  c^\dagger_{\bk 1\sigma}c^\dagga_{\bk+\bQ 2\sigma}    + H.c. \bigg ] \Bigg\},
	\nonumber
\eea
with real hopping parameters $\widetilde{Z}$ given by Eq.~\eqn{Zetas} where each $R_{a\sigma}$  
is substituted by its absolute value $\rho_{a\sigma}$. In addition the new Hamiltonian contains  
spin, orbital and sublattice dependent potential 
\be
V_*(t) = \sum_{\bR\in A, a\sigma}\, 
\dot{\phi}_{a\sigma}(t)\,n_{\bR a \sigma} \,
+ \sum_{\bR\in B, a\sigma}\, 
\dot{\phi}_{a-\sigma}(t)\,n_{\bR a \sigma}.\label{def:V_*}
\ee
It is actually this transformed Hamiltonian $\widetilde{\m{H}}_*(t)$ that provides a more transparent interpretation of the quasiparticle dynamics. In particular, the diagonalization of the instantaneous transformed Hamiltonian may give sensible indications whether the two bands are instantaneously separated by a finite gap. This criterium is however effective only when magnetism is allowed, in which case already the transformed Hamiltonian is able to  describe a non-trivial insulator.  In fact, magnetic order opens additional Bragg gaps at the boundary of the reduced magnetic Brillouin zone, so that an {\sl insulator} can be identified with the case in which the lowest 
band is separated by a finite gap from the next higher one, each band accommodating at most  one electron per site in the reduced zone.

In the paramagnetic case, where the Mott insulator cannot be inferred from spin-symmetry breaking, the above criterium is useless. Moreover, since $R_{a\sigma}(t)$ is complex, it is not even as straightforward as in the single-band case to identify through its temporal evolution a dynamical Mott transition. Nevertheless, there are signals that we believe can be still associated to a dynamical transition. 
We observe that, if we write
\be
\Phi_{\{n\}}(t) = \sqrt{P_{\{n\}}(t)}\;\text{e}^{i\varphi_{\{n\}}(t)},
\ee
where, as we mentioned, $P_{\{n\}}(t)$ is the occupation probability 
of the local Fock state $\ket{n}$ on the correlated wavefunction, 
through Eq.~\eqn{TE2} one realizes that $P_{\{n\}}(t)$ and 
$\varphi_{\{n\}}(t)$ play the role of conjugate dynamical variables. 
In the half-filled single-band model, $\{n\}=(n_1)$, it was found~\cite{Schiro&Fabrizio_prl} that 
the time evolution of the phase 
\be
2\varphi(t) = \varphi_{(0)}(t) + \varphi_{(2)}(t) - 
\sum_{\sigma=\up,\down}\,\varphi_{(\sigma)}(t),
\ee
conjugate to the probability that a site is empty or doubly occupied 
minus the probability that it is singly occupied, i.e. 
$P_{(0)}(t)+P_{(2)}(t) - P_{(\up)}(t)-P_{(\down)}(t)$, 
reflects in a very transparent way the dynamical metal-insulator transition. 
Indeed, in the metallic state $\varphi(t)$ oscillates around a mean 
value, signaling that its conjugate variable is undetermined. On the contrary, in the Mott insulating regime the phase $\varphi(t)$ monotonically increases 
with time; its mean value is thus undetermined unlike the value of its conjugate variable. Since the conjugate variable is nothing but the double occupancy -- at half-filling the probability of a site being empty must be the same as being doubly occupied --  that change of behavior evidently signals the dynamical counterpart of the equilibrium Brinkman-Rice metal-insulator transition,~\cite{Brinkman&Rice} which is how the Mott transition looks like within the Gutzwiller approximation.  

In our two-band model, we shall start from the Mott insulator phase, where the lowest orbital ``1" is Mott localized and the highest ``2" empty, and try to induce a non-equilibrium transition into the two-band metal. It is therefore natural to focus on the same phase variable as before pertaining just 
to orbital ``1", which should turn from being Mott localized to itinerant. In other words, we shall concentrate on the dynamical evolution of the phase 
\be
2\varphi(t) = \varphi_{(0,0)}(t) + \varphi_{(2,0)}(t) - 
\sum_{\sigma=\up,\down}\,\varphi_{(\sigma,0)}(t),\label{phi_tilde}
\ee 
conjugate to the probability that orbital ``1" is empty or doubly occupied minus the probability that is singly occupied, with the orbital ``2" staying empty. More specifically, we shall monitor the time evolution of $\cos\varphi(t)$ with the belief that, if its time-average vanishes, the system is still Mott insulating. On the contrary, if the time-average of $\cos\varphi(t)$ becomes finite we shall conclude that the system has dynamically jumped into the two-band metal regime. 

We just mention that the transformation $\varphi \to \pi+\varphi$ and $c^\dagga_{\bR 1\sigma}
\to  -c^\dagga_{\bR 1\sigma}$ reflects a $Z_2$ gauge symmetry of the Gutzwiller representation~\cite{Schiro&Fabrizio_prb} as well as of the equivalent $Z_2$-slave-spin representation,~\cite{Demedici-Z2,Huber-Z2,Sigrist-Z2} which provides a very simple interpretation of the Mott transition in the enlarged Hilbert space exploited by both techniques. Indeed, the metal phase turns out to correspond to a phase with spontaneous breaking of the global $Z_2$ symmetry, and the Mott transition to the recovery of such symmetry in the Mott insulator.
\cite{Sigrist-Z2,Schiro&Fabrizio_prb,Baruselli-Z2} 

In conclusion, in what follows we shall exploit either the 
time-average of $\cos\varphi(t)$ or the spectrum of the time-averaged Hamiltonian 
$\widetilde{\mathcal{H}}_*(t)$ in Eq.~\eqn{H_*-vera} to establish if the system evolves into a 
metal or insulating state. As discussed, the choice will depend whether magnetism survives at long times.

\begin{figure}[h]
\includegraphics[scale=0.32]{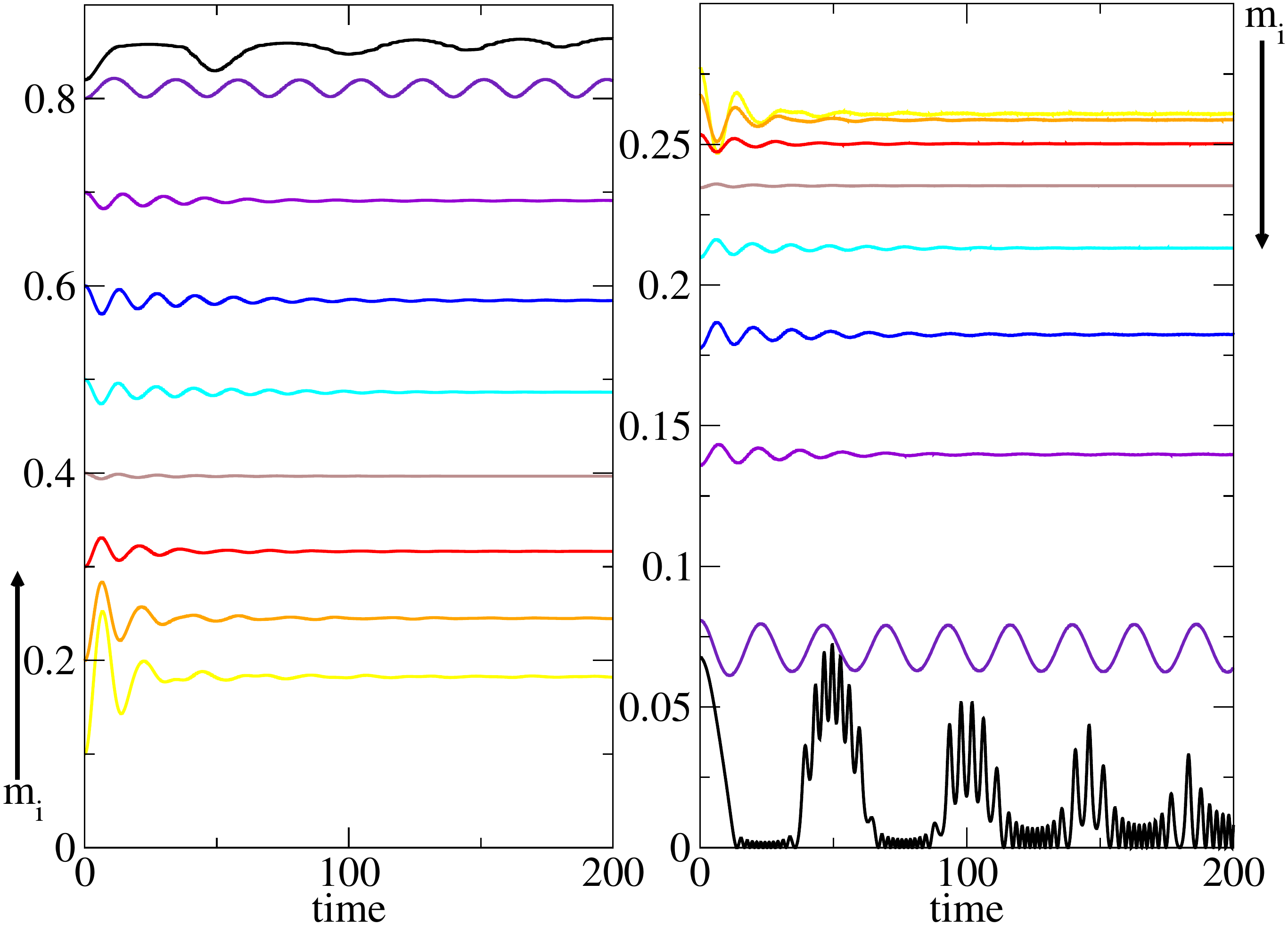}
\caption{(Color online) Time evolution of $m(t)$ (left panel) and 
of the orbital ``1'' renormalization factor $|R_1|^2(t)$ (right panel) for different values of $m_i$. In the left panel the order of the curves from the bottom is $m_i=0.1,0.2,\dots,0.8,0.82$, while on the right panel the same order is from the top. In this as well as in all other figures time is measured in units of $8/W$. }
\label{fig:MvsT}
\end{figure}

\section{Non-equilibrium dynamics in the Gutzwiller approximation}
\label{sec5}
Our aim is to simulate the out-of-equilibrium process described in 
section \ref{The model out-of-equilibrium} within the time-dependent Gutzwiller approximation. The first issue is how to initialize the state after the fast impulse has transferred electrons from the lower orbital to the upper one. We make here the {\sl adiabatic} assumption that such initial state is the lowest energy one at fixed orbital polarization $m_i< m^\text{eq}$, where $m^\text{eq}$ is the equilibrium value corresponding to the Hamiltonian parameters.    
This assumption realizes just the processes depicted in Figs.~\ref{EnevsM_PM} and \ref{EnevsM_AFM}, where the system is instantaneously endowed with a value $m_i$ of the orbital polarization. Such initial state is then let evolve according to the equations \eqn{TE1} and \eqn{TE2}.

We shall consider separately the case in which the system is forced to evolve in the paramagnetic sector and the more realistic one in which magnetism is allowed. 
 
\subsection{Paramagnetic dynamics}

We assume Hamiltonian parameters such that the system at equilibrium and at $T=0$ is a one-band Mott insulator not far from the transition to a two-band metal phase. Specifically we take $U=1.875W$ and $2\Delta=0.025W$ (red bullets in Fig.~\ref{fig:phaseT0}), so that 
at equilibrium $m^\text{eq}=1$.  As mentioned, we initialize the state with an initial orbital polarization 
$m_i<m^\text{eq}$ and study the time evolution for different $m_i$'s. 
 
\begin{figure}[th]
\includegraphics[scale=0.32]{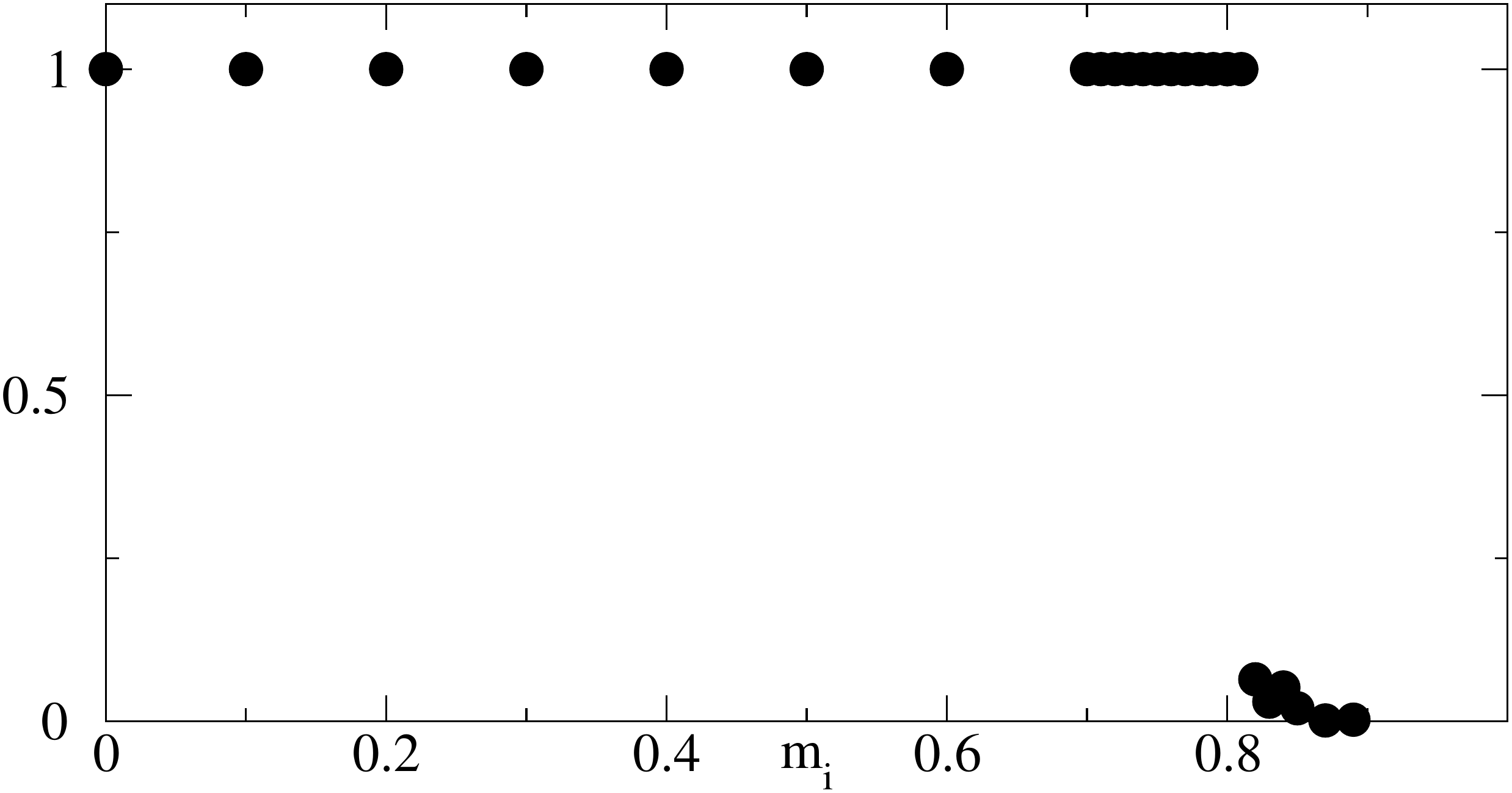}
\caption{Time-average of $\cos\varphi(t)$ as function of the initial value of the orbital polarization $m_i$. }
\label{cos-phi}
\end{figure}
In Fig.~\ref{fig:MvsT} we show the time evolution of the orbital polarization, $m(t)$, and of the 
quasiparticle residue of the orbital``1'', $|R_1(t)|^2$, for different initial $m_i$'s. We readily recognize two distinct dynamical regimes. For small values of $m_i$, i.e. far from the equilibrium value $m^\text{eq}=1$,  both the orbital polarization and $|R_1|^2(t)$ show a damped oscillatory evolution towards steady-state values. 
We can actually distinguish two well separate sets of oscillation frequencies, the shorter one being almost invisible in the figure. 
We observe that $m_i\simeq 0.3$ corresponds to the metastable minimum in Fig. \ref{EnevsM_PM}, which, being a saddle point of the energy functional $E[m]$, is stationary with respect to the Gutzwiller equations of motion. Even though we cannot establish a strict correspondence, still the 
dynamics of the orbital polarization $m(t)$ follows that of a classical particle in a potential $E[m]$. 
This foresees that as $m_i$ increases (less energy is injected into the system) and crosses over 
the top of the barrier separating the metallic relative minimum from the insulating absolute one, the time evolution radically changes. In this case, in fact, we 
do not find anymore relaxation to a steady-state, see for example the case $m_i = 0.8$ in Fig.~\ref{fig:MvsT}, and an undamped oscillating mode persists. Finally, when
$m_i$ is further increased,  see the case $m_i = 0.82$ in Fig.~\ref{fig:MvsT}, the dynamics changes abruptly: $|R_1|^2$ approaches zero and the
faster oscillations become more and more visible. The orbital polarization $m(t)$ displays a less regular behavior due to the small values of the renormalized hopping parameters that freeze the dynamics of the Slater determinant. Overall, $m(t)$ does not display significant deviations from its initial value.

More information can be gained by the behavior of the phase variable defined in Eq.~\eqn{phi_tilde}, 
more specifically of the long-time average of $\cos\varphi(t)$, which is shown in Fig. \ref{cos-phi}.
We observe that the time average is essentially vanishing for large $m_i$ but,  below 
$m_{*}\simeq 0.8$, abruptly jumps to a finite value close to one. As discussed earlier, we take this as 
signature of a dynamical phase transition from the Mott insulator at $m_i\geq m_*$ to the 
two-band metal at $m_i< m_*$. The evidence that, for $m_i\geq m_*$, $m(t)$ oscillates around a finite value, indicates that the out-of-equilibrium Mott insulator has to be regarded as an excited state 
with electrons in the conduction band and holes in the valence one. 

\begin{figure}[!h]
\includegraphics[scale=0.32]{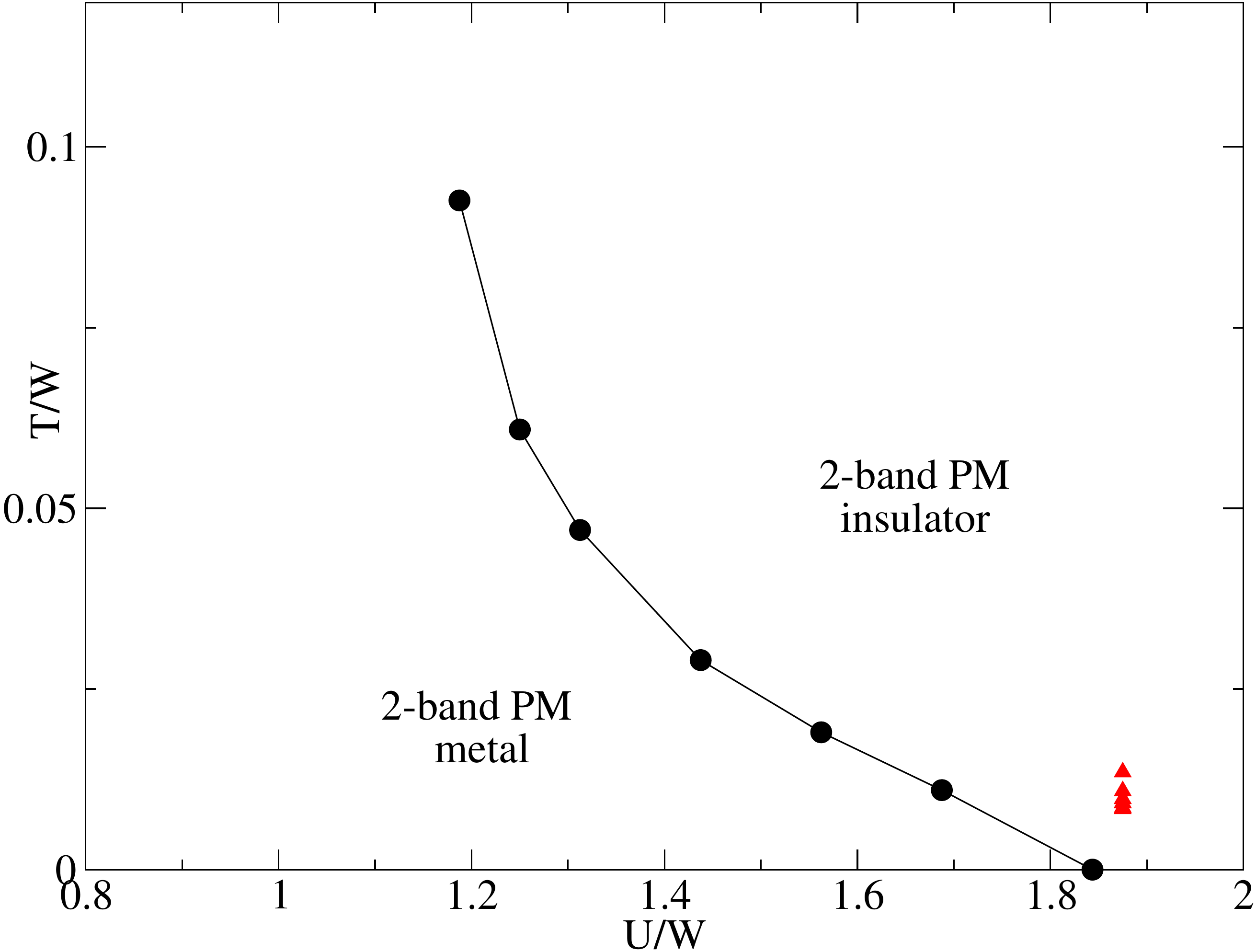}
\caption{(Color online) Phase diagram temperature $T$ versus $U$ at $\Delta=0.025 W$ within the paramagnetic sector. 
The red triangles indicate the effective temperatures at which our non-equilibrium state would correspond 
if thermalization holds, for values of $m_i=0.1,\dots,0.8$.}
\label{Teff}
\end{figure}

We highlight that such a metal regime is not compatible with the hypothesis that the energy supplied to the system simply heats it. Indeed, if we transform, following the thermalization hypothesis, this excess energy into a temperature determined by imposing that the total energy, conserved in the unitary evolution, coincides with the internal energy at that temperature, we obtain the points shown in Fig.~\ref{Teff}, all of which 
are inside the Mott insulating phase. 

In other words, the metal regime that seems to be stabilized during the dynamical evolution is incompatible with thermalization, but it is rather related to the metastable metallic minimum shown in Fig.~\ref{EnevsM_PM}. Therefore the evidences seem to confirm the expectations of section~\ref{The model out-of-equilibrium} that, when the Mott transition is first order, it is possible to stabilize a metastable metal by properly driving off-equilibrium the Mott insulator.

\subsection{AFM dynamics}
\label{AFM dynamics}

We now repeat the same analysis without enforcing paramagnetism. Specifically, we shall focus here on the time evolution of an initial correlated AFM state subject to a sudden redistribution of the orbital polarization, as sketched in Fig. \ref{EnevsM_AFM}. 

In Fig.~\ref{EnevsM_AFM} we have shown the energy $E[m]$ as a function of the orbital polarization for $U=1.125W$. We note that a PM metallic minimum and an AFI one coexist, with their respective energies that cross as a function of the crystal field. We observe that the insulating solution is not fully polarized, $m_{eq}\neq1$, since the AFM insulator within the GA has finite hopping renormalization factors, so that the inter-orbital hybridization is finite.  
We do expect that also in this case a stable paramagnetic metal can emerge without any thermal counterpart.  

We thus generalize the orbital polarization quench of the previous Section to study the evolution of an initial $T=0$ AFM state at the fixed values of $2\Delta=0.05W$ and $U=1.125W$, $1.375W$ and $1.625W$ (red bullets in Fig.~\ref{fig:phaseT0}). If we denote with $n^\text{eq}_{a \sigma}$ the equilibrium occupation 
at sublattice $A$ of orbital $a$ with spin $\sigma$, the initial nonequilibrium state is built by minimizing the Gutzwiller energy imposing that 
\bea
n^i_{1\sigma} &=& \alpha \; n^\text{eq}_{1\sigma} \nonumber \\
n^i_{2\sigma} &=& n^\text{eq}_{2\sigma} + (1 - \alpha) \; n^\text{eq}_{1\sigma} \;, 
\eea
with $\alpha<1$ quantifying the deviation from the equilibrium value $\alpha^\text{eq}=1$.  
The state so constructed mimics an initial excited configuration in which electrons 
are transferred from the lower band to the upper one without flipping their spin, thus 
leaving unaltered the staggered magnetization. 
\begin{figure}[!h]
\includegraphics[scale=0.32]{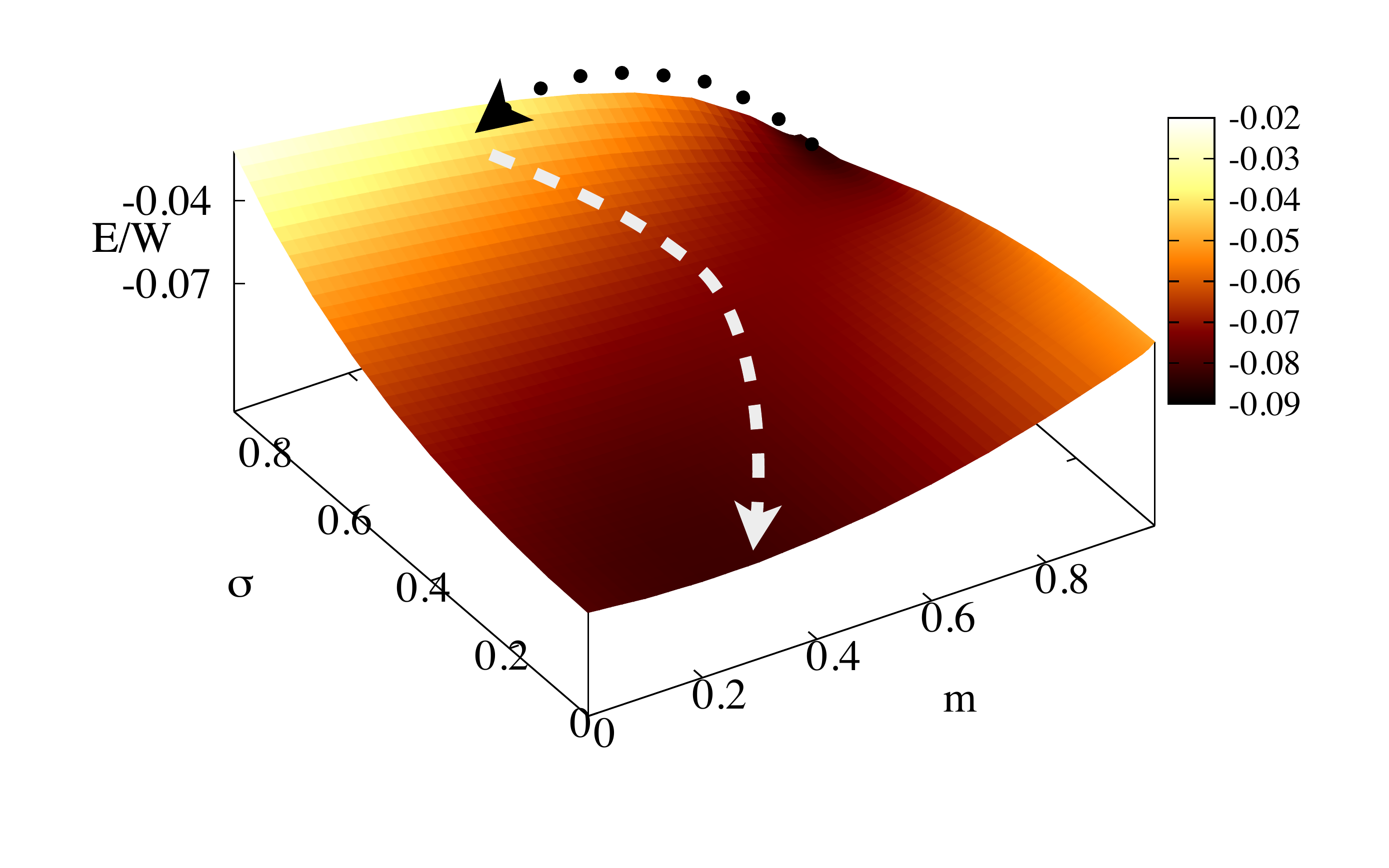}
\caption{(Color online) Energy functional at fixed value of the orbital polarization and of the total magnetization for a value of $2\Delta=0.05W$. The black dotted line represents the quench in the orbital polarization, while the grey line sketches an approximate classical dynamics.}
\label{fig:surface}
\end{figure}
In other words, the initial magnetization does not correspond to the optimized one at the given value of the initial orbital polarization $m_i$. This is evident in Fig. ~\ref{fig:surface} where we plot the energy functional $E\big[m,\sigma]$, where both orbital polarization $m$ and staggered magnetization 
$\sigma$ are fixed. The two minima in the Fig.~\ref{EnevsM_AFM} correspond respectively to a paramagnetic metallic state and to an antiferromagnetic insulating one, the latter being the actual absolute minimum. We remark the absence of a third minimum corresponding to a paramagnetic Mott insulator, which is unstable at $T=0$.

\begin{figure}[!h]
\includegraphics[scale=0.32]{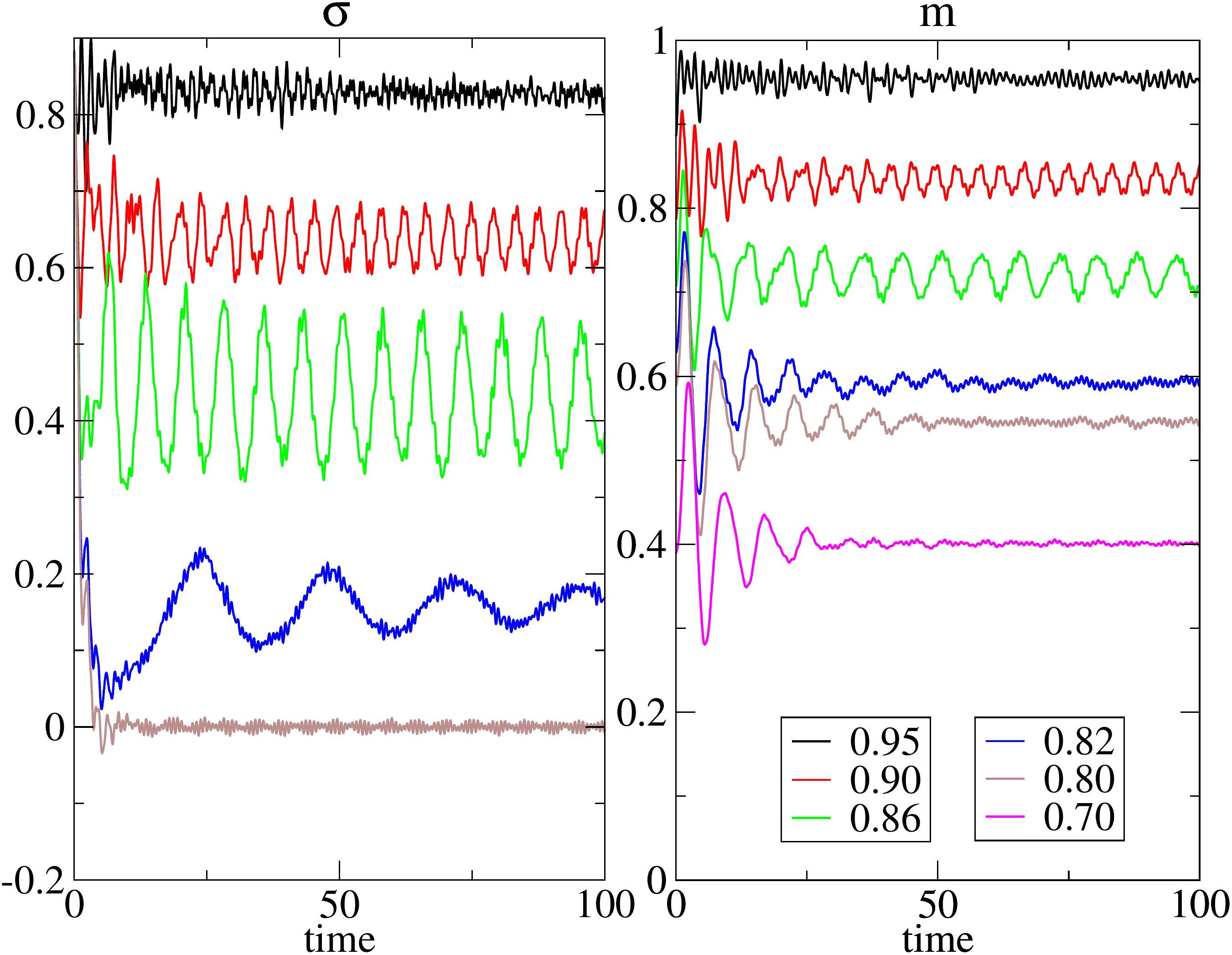}
\includegraphics[scale=0.32]{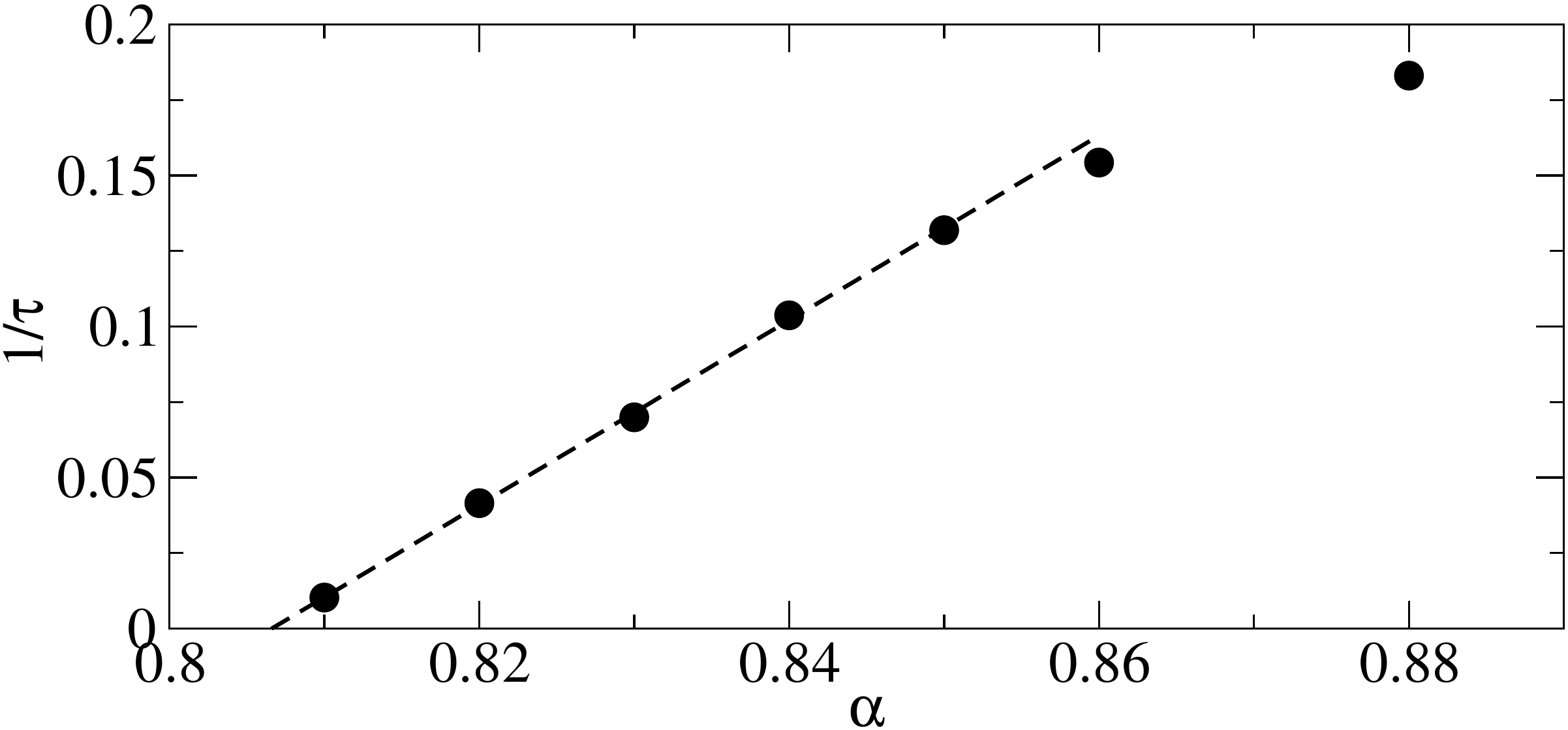}
\caption{(Color online) Upper panels: Time evolution of the staggered magnetization $\sigma(t)$ (left panel) and of the orbital polarization $m(t)$ (right panel) for different values of $\alpha$ at a fixed value of $U=1.125W$. Lower panel: Inverse of the period oscillation for the AFM coherent mode as a function of the quench parameter $\alpha$.}
\label{fig:U9}
\end{figure}

From the figure one can imagine that upon quenching the orbital polarization 
beyond the barrier, black dotted arrow in the figure, $m(t)$ and $\sigma(t)$ will be attracted towards the metastable paramagnetic metallic minimum, grey arrow in the figure. Of course this argument is just qualitative since the energy surface is self-consistently coupled to the dynamics of all other degrees of freedom besides $m$,  hence it is not constant in time. 
Nevertheless, we find that this na\"ive  expectation is qualitatively correct. In Fig.~\ref{fig:U9} we show the evolution of the orbital polarization $m(t)$ and of the staggered magnetization $\sigma(t)$ for different $\alpha$'s at the smallest value of $U=1.125W$. We first note that, as in the paramagnetic case, a high frequency oscillating pattern superimposes on top of a much slower oscillation. Moreover, a further frequency scale exists and it is associated to the magnetic order, as evident in the evolution of $\sigma(t)$. Upon decreasing $\alpha$, i.e. moving away from equilibrium, $\sigma(t)$ shows indeed a coherent oscillating mode with an increasing period that finally diverges around $\alpha\simeq 0.81$, above which the 
order parameter relaxes to zero. We estimate the frequency associated with magnetic order by the inverse time-distance between the first two maxima in the oscillations of 
$\sigma(t)$, which is plotted in the bottom panel of Fig.~\ref{fig:U9} and vanishes linearly at the transition to the PM phase.  

\begin{figure}[!h]
\includegraphics[scale=0.32]{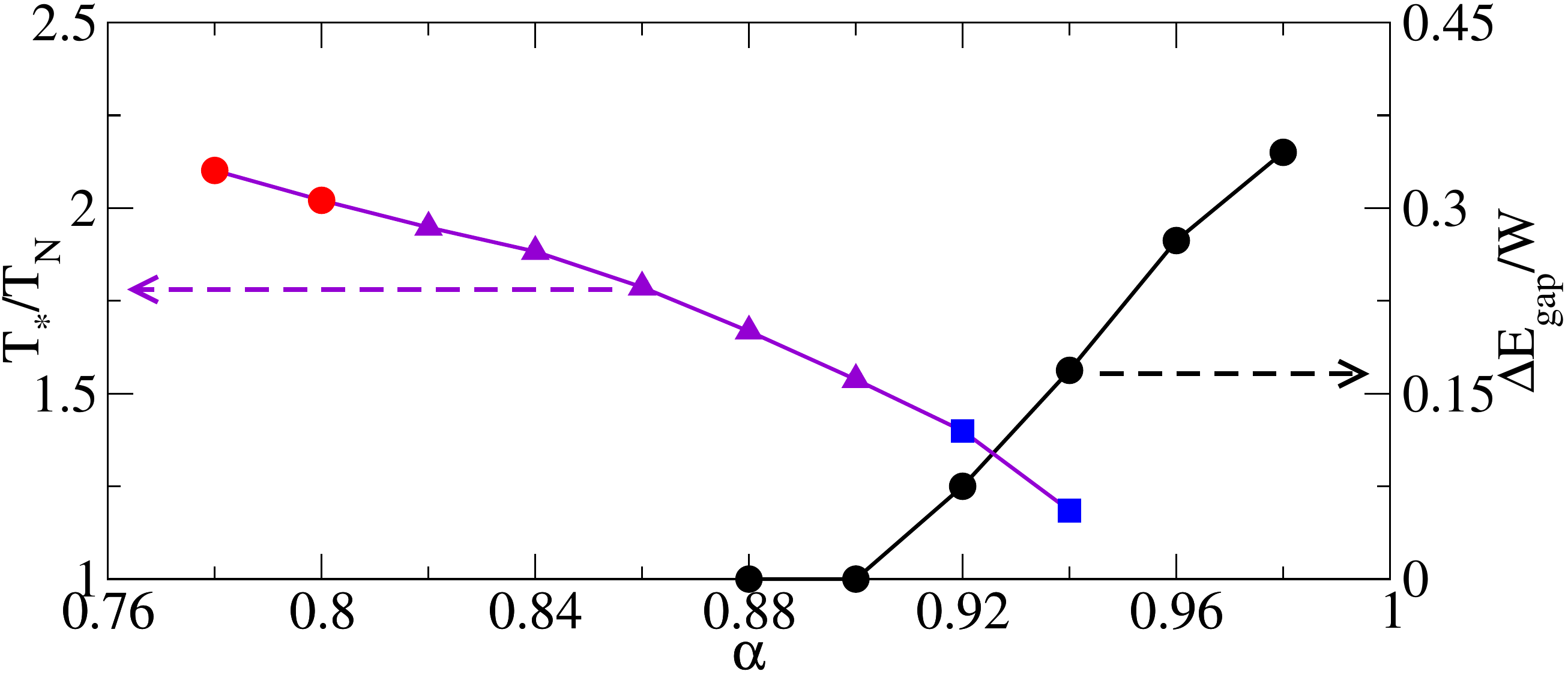}
\caption{(Color online) Leftmost points: Effective temperature $T_*$ in units of the N\'eel temperature $T_N$ extracted 
according to thermalization hypothesis. The blue square points indicate that a magnetic insulator, the purple triangles a SDW metal, and finally the red circles a PM metal. Rightmost points: Energy gap 
$\Delta E_\text{gap}$ between the two lowest bands. Both $T_*$ and $\Delta E_\text{gap}$ are plotted versus $\alpha$. We recall that $\alpha^\text{eq}=1$ is the equilibrium value, so the smaller $\alpha$ the greater the deviation from equilibrium is.}
\label{fig:gap}
\end{figure}

In order to better characterize the state towards which the system flows, we have diagonalized the long-time limit of the Hamiltonian $\widetilde{\mathcal{H}}_*(t)$ in Eq.~\eqn{H_*-vera} and calculated the gap 
$\Delta E_\text{gap}$ between the two lowest bands, the lower one having predominantly the character of orbital ``1" and the upper of orbital ``2". $\Delta E_\text{gap}$ versus $\alpha$ is shown in 
Fig.~\ref{fig:gap}. We observe that the gap closes, i.e. the system turns metallic,  below $\alpha\sim 0.9$, but there is a region $0.81 \lesssim \alpha \lesssim 0.9$ when the magnetic order parameter 
is still finite; this state is therefore a SDW metal. Only below $\alpha\simeq 0.81$ the magnetic order melts and the metal becomes paramagnetic. The existence of a SDW metal is unexpected since such a phase does not appear in the phase diagram Fig.~\ref{fig:phaseFT} at high temperature, which is suggestive of a non-thermal behavior. Indeed, if we extract the effective temperature $T_*$ that would correspond to the initial non-equilibrium condition according to thermalization, also shown in Fig.~\ref{fig:gap}, we observe that magnetic order survives well above a temperature corresponding to the equilibrium N\'eel temperature $T_N$. Even though the Gutzwiller approximation presumably overestimates the effective temperature that corresponds to a given internal energy, we mention that the non-thermal persistence of magnetism has been also observed in the dynamics following an interaction quench of an initial AFM state for the single band Hubbard model.~\cite{Werner_1, Werner_2,Sandri_AFM}

 \begin{figure}[!th]
\includegraphics[scale=0.32]{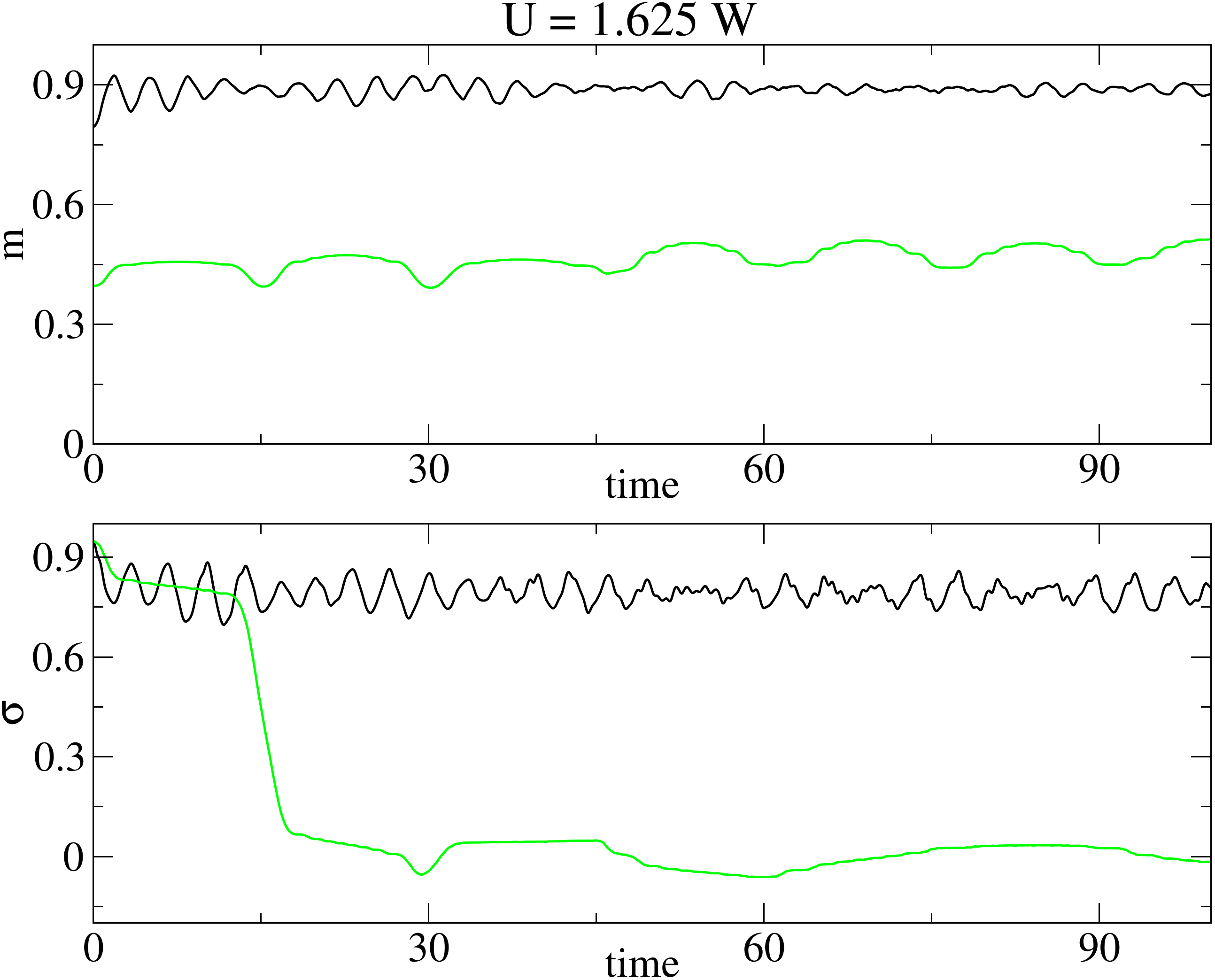}
\caption{(Color online) Time-evolution of the orbital polarization $m(t)$ (upper panel) and of the magnetic order parameter $\sigma(t)$ (lower panel) for two different $alpha$'s at $U=1.625W$. The upper black curves in both panels correspond to 
$\alpha=0.9$, and the green lower ones to $\alpha=0.7$. }
\label{fig:U2_m}
\end{figure}
Apart from the non-thermal persistence of magnetic order, the overall dynamical behavior at 
$U=1.125W$ is what we would expect; above a threshold value of the injected energy the antiferromagnetic Mott insulator turns into a metal as if temperature raises, see the arrow at $U=1.125W$ in Fig.~\ref{fig:phaseFT}. 

\begin{figure}[!h]
\includegraphics[scale=0.32]{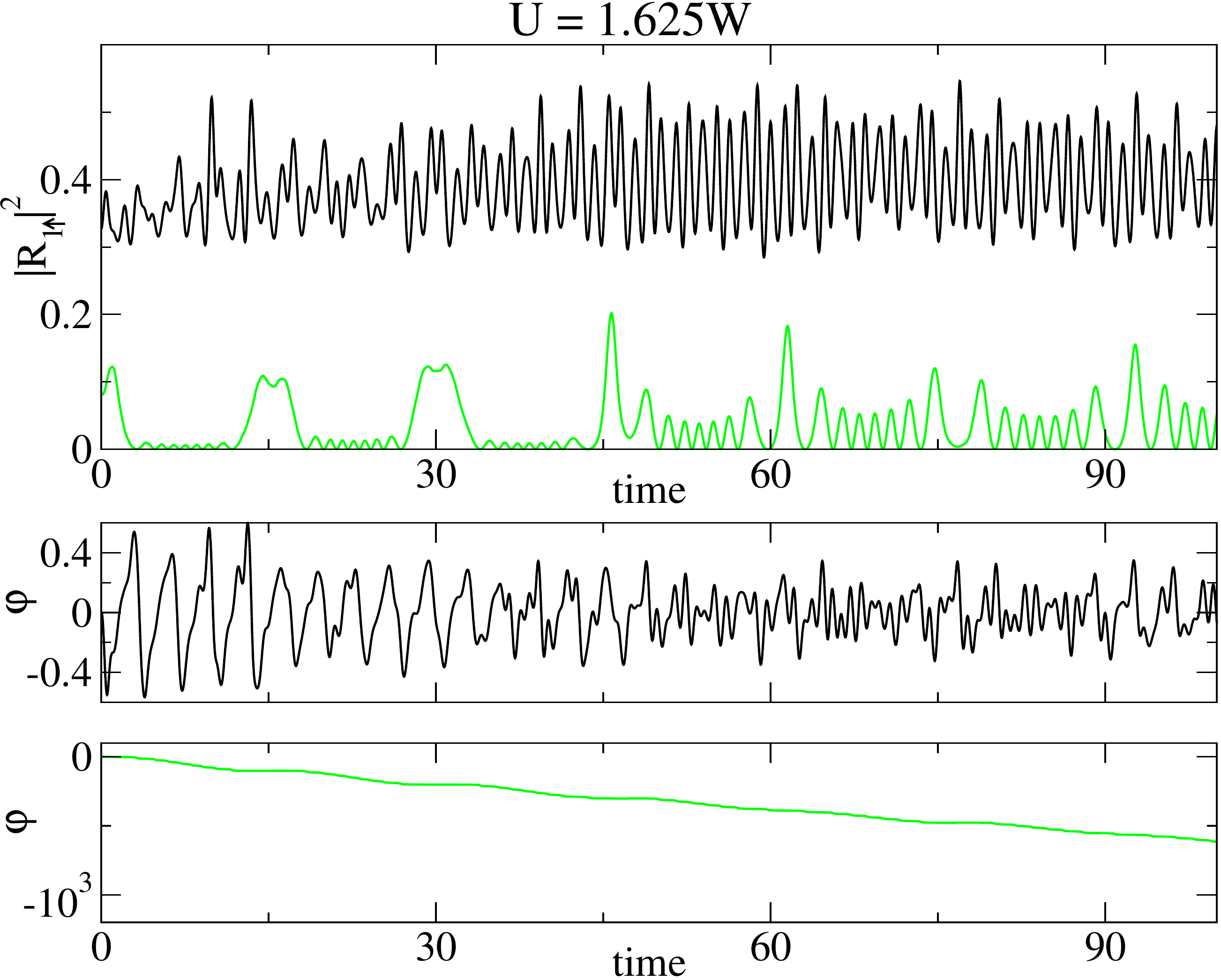}
\caption{(Color online) Top panel: time evolution of the renormalization factor $|R_{1\up}(t)|^2$ for orbital ``1" and majority spin for $\alpha=0.9$, black top curve, and $\alpha=0.7$, green bottom one, at $U=1.625$. 
Bottom panel: phase angle $\varphi(t)$, Eq.~\eqn{phi_tilde}, for $\alpha=0.9$, black curve in the top panel, and $\alpha=0.7$, green one in the bottom panel, still at $U=1.625W$. }
\label{fig:U2_R}
\end{figure}

Seemingly, we would expect that at large $U$ the antiferromagnetic Mott insulator 
should instead transform into a paramagnetic Mott insulator, see again Fig.~\ref{fig:phaseFT}. 
This is indeed what happens at $U=1.625W$. In Fig.~\ref{fig:U2_m} we show the time evolution of the 
orbital polarization $m(t)$ and of the magnetic order parameter $\sigma(t)$ for two different $alpha$'s, 
$\alpha=0.9$ close to equilibrium and $\alpha=0.7$ further from it. We observe that, while at $\alpha=0.9$ 
the magnetic order parameter stays finite, at $\alpha=0.7$ it flows to zero. The spectrum of the 
effective Hamiltonian $\widetilde{\mathcal{H}}_*(t)$, Eq.~\eqn{H_*-vera}, at large times shows that 
$\alpha=0.9$ still corresponds to a magnetic Mott insulator, with a well defined gap. In the other case, 
$\alpha=0.7$, we have instead to resort to the phase variable Eq.~\eqn{phi_tilde} to establish whether the 
paramagnetic state is metallic or insulating. In Fig.~\ref{fig:U2_R} we thus show the time evolution of the renormalization factor $|R_{1\up}(t)|^2$ for orbital ``1" and majority spin and of the phase angle 
$\varphi(t)$ defined in Eq.~\eqn{phi_tilde}. We observe that $\varphi(t)$ at $\alpha=0.7$ decreases monotonically with time, which we take as indication that the paramagnetic state towards which the system flows is indeed insulating. We highlight the different dynamical behavior of the  $\alpha=0.9$ magnetic insulator, where $|R_{1\up}(t)|^2$ oscillates around a finite value and $|\varphi(t)|$ does not grow indefinitely.  

\begin{figure}[!bh]
\includegraphics[scale=0.32]{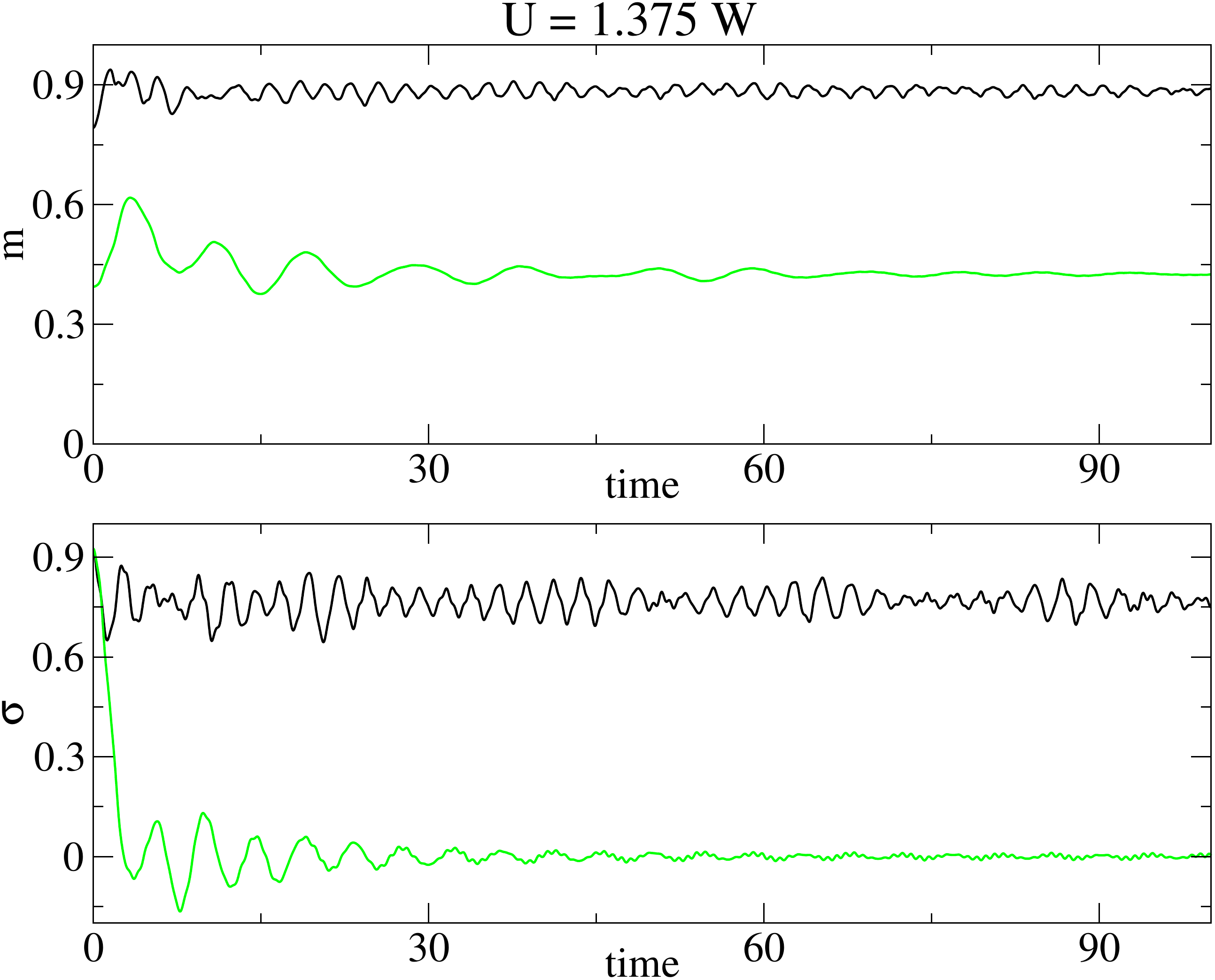}
\caption{(Color online) Same as Fig.~\ref{fig:U2_m} but at $U=1.375W$.}
\label{fig:U1_m}
\end{figure}
Therefore, both at small $U=1.125W$ and large $U=1.625W$ the non-equilibrium pathway is essentially 
equivalent to a temperature rise. Quite different is instead the case when $U$ is closer to the 
paramagnetic metal to paramagnetic insulator first order line. Fig.~\ref{fig:U1_m} is the same as 
Fig.~\ref{fig:U2_m} but at smaller $U=1.375W$, see Fig.~\ref{fig:phaseFT}. We observe that, as expected, 
upon decreasing $\alpha$, i.e. deviating more from equilibrium, the magnetic order parameter disappears. 
However, unlike the case at  $U=1.625W$, the phase angle $\varphi(t)$ remains limited in a finite window also when magnetism melts. From the spectrum of the 
effective Hamiltonian $\widetilde{\mathcal{H}}_*(t)$, Eq.~\eqn{H_*-vera}, we also deduce that the 
gap is finite at $\alpha=0.9$ but vanishes at $\alpha=0.7$. This result together with the behavior of the 
phase angle $\varphi(t)$ suggests that upon moving away from equilibrium, i.e. reducing $\alpha$, the 
antiferromagnetic Mott insulator eventually gives up to a paramagnetic metal, which has no counterpart in the equilibrium phase diagram, see Fig.~\ref{fig:phaseFT}. At first glance this finding might look odd. 
However, we earlier remarked that close to the crossing point between the N\'eel critical line and the first order line separating at $T>T_N$ the paramagnetic metal from the paramagnetic insulator, there actually exist three coexisting phases. Therefore, even though at $U=1.375W$ the equilibrium phase diagram only 
displays either an antiferromagnetic insulator or a paramagnetic one, a metastable paramagnetic metal does exist and, as we just showed, can be accessed by a non-equilibrium pathway. 
\begin{figure}[th]
\includegraphics[scale=0.32]{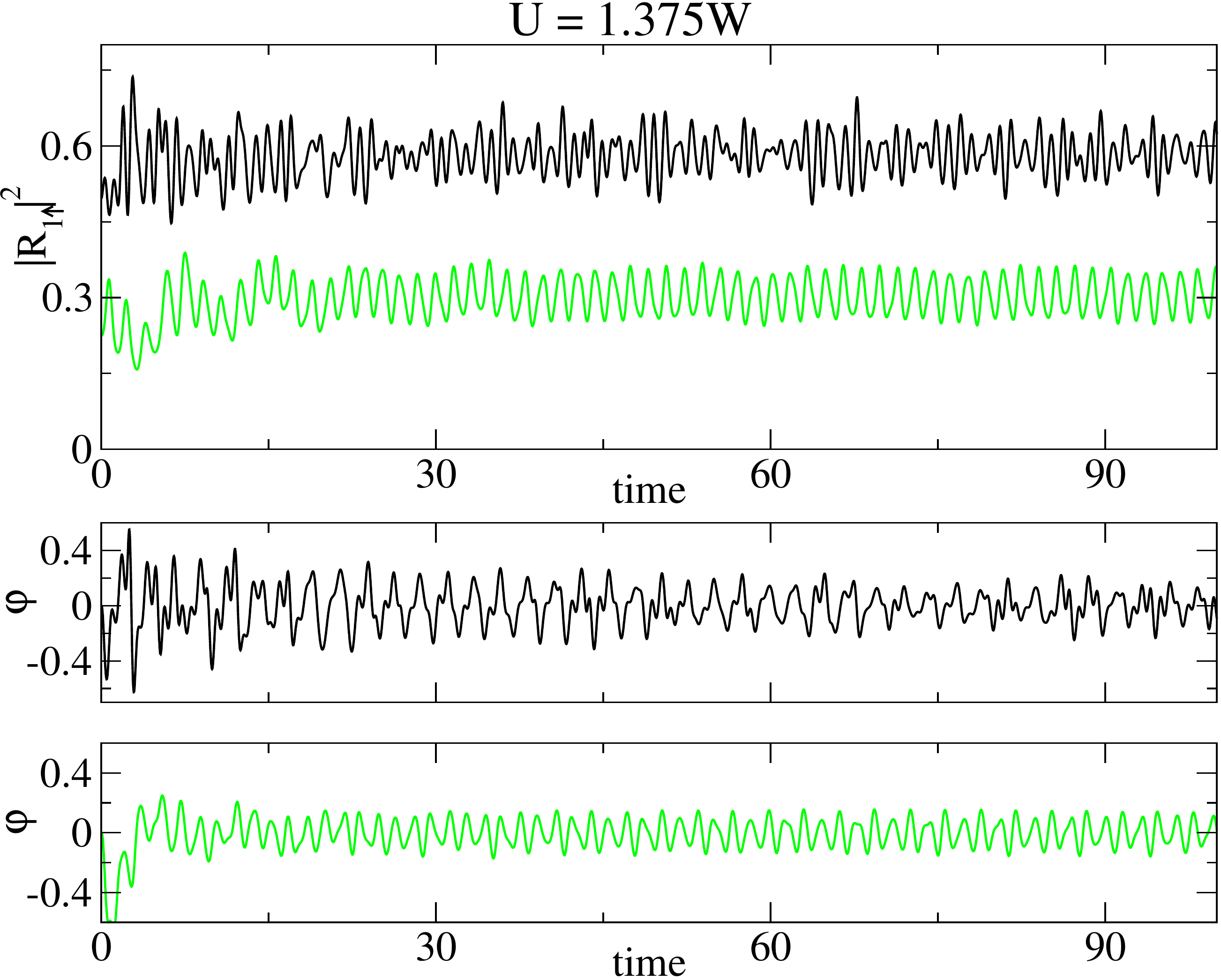}
\caption{(Color online) Same as Fig.~\ref{fig:U2_R} but at $U=1.375W$.}
\label{fig:U1_R}
\end{figure}


\section{Conclusions}
\label{Conclusions}
We have studied the out-of-equilibrium dynamics of a quarter-filled Hubbard model of two 
orbitals that are weakly crystal-field split, which we believe captures the key physical properties of vanadium sesquioxide. 

We showed that exciting the Mott insulator by suddenly transferring electrons from the lower orbital to the upper one may lead to a gap-collapse and drive the system into a metastable metal phase. Such a peculiar non-equilibrium pathway has been uncovered by means of the time-dependent Gutzwiller approximation. Even though this method suffers from a lack of dissipative channels that operate in the real dynamics, hence clearly overestimates the time the system stays trapped into the metastable metal, nevertheless we believe that the qualitative scenario is correct. The reason is that it simply derives, and actually could have been predicted in advance, from general properties at equilibrium that have been also found by more rigorous DMFT treatments.~\cite{Ferrero-2band,Sandri_finT} 
Specifically: 
\begin{itemize}
\item[(1)] The Mott insulator, 
either paramagnetic or antiferromagnetic, is characterized by a minimal gap that corresponds to transferring 
electrons from the occupied lower orbital to the empty upper one, rather than from the lower to the upper Hubbard sub-bands. Moreover that gap is a dynamical, breathing quantity, in the sense that it determines but it is in turns determined by the relative occupation of the two orbitals;  
\item[(2)] The Mott transition is generically first order even at zero temperature. 
\end{itemize}
In fact, point (1) implies that a non-equilibrium excess population of the upper 
orbital might lead to a gap closing and temporarily push the system in the metastable metal phase whose 
existence on the insulating side nearby the transition is entailed by point (2).

Although the model is a very specific one, we are convinced that the overall physical behavior might be applicable to the class of Mott insulating materials that display the two characteristic properties (1) and (2) above. Such a class presumably includes V$_2$O$_3$ and VO$_2$, and possibly also some charge transfer Mott insulators. As a matter of fact, non-equilibrium gap collapse has been indeed  observed recently by pump-probe time-resolved photoemission in V$_2$O$_3$,~\cite{Marino} and also in 
VO$_2$,~\cite{Rubio} even though 
it is still under debate the relevance of strong correlations in the insulating phase of 
VO$_2$.~\cite{Georges-VO2,Eyert} In addition, the important role of the first order character of the Mott transition that we unveiled is consistent with the phenomenological model introduced in Ref.~\onlinecite{Cario-2013} to interpret the dielectric breakdown observed in several Mott insulators.

\begin{acknowledgments}
We acknowledge useful discussions with E. Tosatti, A. Amaricci and G. Mazza. We thank G. Lantz and M. Marsi for kindly showing us  their V$_2$O$_3$ time-resolved ARPES data included in Ref.~\onlinecite{Marino} prior to their publication. 
This work has been supported by the European Union, Seventh Framework Programme, under the project GO FAST, Grant Agreement no. 280555. 
\end{acknowledgments}


\end{document}